\documentclass[useAMS,usenatbib]{mn2e}
  \usepackage{times}
  \usepackage[usenames]{color}
  \usepackage{graphicx}
  \usepackage{url}
  \usepackage{hyperref}
  \usepackage{breakurl}

\title[The \xmm-LSS Catalogue II]
      {The \xmm-LSS catalogue: X-ray sources and associated multiwavelength data. Version II}
\author[L.Chiappetti et al]
      {L.~Chiappetti$^{1}$\thanks{E-mail: lucio@lambrate.inaf.it},
       N.~Clerc$^{2,5}$, F.~Pacaud$^{3}$, M.~Pierre$^{2}$,
       \newauthor
       A.~Gu\'eguen$^{2,6}$,
       L.~Paioro$^{1}$,
       M.~Polletta$^{1}$,
       O.~Melnyk$^{4,7}$, A.~Elyiv$^{4,8}$, J.~Surdej$^{4}$,
       L.Faccioli$^{2}$
       \\
       $^1$INAF, IASF Milano, via Bassini 15, I-20133 Milano, Italy\\
       $^2$Laboratoire AIM, CEA/ DSM/Irfu/SAp, CEA-Saclay, F-91191
           Gif-sur-Yvette C\'edex, France\\
       $^3$Argelander Institut f\"ur Astronomie, Universit\"at Bonn,
           Auf dem H\"ugel 71, D-53121 Bonn, Germany\\
       $^4$Institut d'Astrophysique et de G\'eophysique, Universit\'e de Li\`ege,
           All\'ee du 6 Ao\^ut,   17, B5C, 4000 Sart Tilman, Belgium\\
       $^5$present address: Max-Planck-Institut f\"ur extraterrestrische Physik,
           Giessenbachstra{\ss}e, D-85748 Garching, Germany\\
       $^6$present address: GEPI, Observatoire de Paris, CNRS, Universit\'e Paris Diderot,
           5, place Jules Janssen, F-92195 Meudon, France\\
       $^7$ Astronomical Observatory, Kyiv National University, vul. Observatorna 3,
           04053 Kyiv, Ukraine\\
       $^8$ Main Astronomical Observatory, Academy of Sciences of Ukraine,
           vul. Akademika Zabolotnoho 27, 03680 Kyiv, Ukraine\\
       }

\begin{document}

 \newcommand{\xmm}{\textit{XMM}}
 \newcommand{\xmmn}{\textit{XMM-Newton}}
 \newcommand{\xlssII}{{\tt 2XLSS}}
 \newcommand{\xlssd}{{\tt 2XLSSd}}
 \newcommand{\xlss}{{\tt XLSS}}
 \newcommand{\xamin}    {{\sc Xamin}}
 \newcommand{\dd}{$\rm deg^{2}$}
 \newcommand{\flux}{$\rm erg \, s^{-1} \, cm^{-2}$}
 \renewcommand{\UrlBreaks}{\do\/\do\a\do\b\do\c\do\d\do\e\do\f\do\g\do\h\do\i\do\j\do\k\do\l\do\m\do\n\do\o\do\p\do\q\do\r\do\s\do\t\do\u\do\v\do\w\do\x\do\y\do\z\do\A\do\B\do\C\do\D\do\E\do\F\do\G\do\H\do\I\do\J\do\K\do\L\do\M\do\N\do\O\do\P\do\Q\do\R\do\S\do\T\do\U\do\V\do\W\do\X\do\Y\do\Z}

 \date{revised version submitted 2012 October 12; in original form 2012 June 22}
 \pagerange{\pageref{firstpage}--\pageref{lastpage}} \pubyear{2012}
 \maketitle
 \label{firstpage}

 \begin{abstract}
We present the final release of the multi-wavelength \xmm-LSS data set,
covering the full survey area of 11.1 square degrees, with X-ray data
processed with the latest \xmm-LSS pipeline version. The present
publication supersedes the \cite{2007MNRAS.382..279P} catalogue pertaining to
the initial 5 \dd. We provide X-ray source lists in the customary
energy bands (0.5-2 and 2-10 keV) for a total of 6721 objects in the deep 
full-exposure catalogue and 5572 in the 10ks-limited one, above a 
detection likelihood of 15 in at least one band. We also provide a
multiwavelength catalogue, cross-correlating our list with IR, NIR, 
optical and UV catalogues. Customary data products (X-ray FITS images, 
CFHTLS and SWIRE thumbnail images) are made available together with our 
interactively queriable database in Milan, while a static snapshot of the 
catalogues has been 
 supplied to CDS. 
 \end{abstract}


\section{Introduction}
\label{SecIntro}

The rationale for the \xmm-Large Scale Structure (\xmm-LSS) survey was presented
in \cite{2004JCAP...09..011P}. A first catalogue for the 5.5 \dd~ surveyed until  year
2003 was presented in \citet[hereafter Paper I]{2007MNRAS.382..279P}.
In the present paper, we supersede the first release with a new
version which covers the entire 11.1 \dd area of the survey. All the
data were processed or re-processed afresh with the latest version of our pipeline
(see Section \ref{SecXamin}).
We release two families of X-ray database tables 
(see Section \ref{SecOnline}),
a standard catalogue (termed \xlssII) for event files truncated to a common uniform exposure
of 10 ks, and a \textit{deeper} catalogue (termed \xlssd) using
the full exposure time.

The \xmm-LSS survey area, located around $2^h30^m$~ $-5\degr$, was covered
in the optical band by the Canada France Hawaii Telescope Legacy 
Survey\footnote{\url{http://cfht.hawaii.edu/Science/CFHTLS/}}
Wide and Deep Synoptic fields (CFHTLS-W1 and D1);
in the NIR band partially by the UKIRT Infrared Deep Sky 
Survey\footnote{\url{http://www.ukidss.org/}}
(UKIDSS; \citealt{2007MNRAS.379.1599L});
in the IR by the 
\textit{Spitzer} Wide-area InfraRed Extragalactic 
survey\footnote{\url{http://swire.ipac.caltech.edu/swire/swire.html}}
(SWIRE; \citealt{2003PASP..115..897L});
and in the UV by the Galaxy Evolution Explorer\footnote{\url{http://www.galex.caltech.edu/}}
(\textit{GALEX}; \citealt{2005ApJ...619L...1M}) all-sky survey.
We release also a multiwavelength database table (using data from the sources
just described) in conjunction with each of the X-ray table families.

Data from the present catalogue have already been used in other works, published,
submitted or in preparation,
e.g. \cite{2011A&A...526A..18A}, \cite{2012A&A...537A.131E},
\cite{2012Willisinprep},
\cite{2012Melnykinprep},
\cite{2012Clercinprep}.

 The plan of the paper is as follows:
 in Section \ref{SecNEWCat} we describe the layout and content of our catalogue, in
 particular Section \ref{SecOnline} presents our database system by which users can have
 public access to the entire data tables and associated data products
 (a reduced summary will be available via
 the Centre de Donn\'ees de Strasbourg [CDS\footnote{\url{http://cdsweb.u-strasbg.fr}}]);
 Section \ref{SecNEWProc} describes the X-ray data processing and
 Section \ref{SecOpt} the generation of the multiwavelength catalogue.
 Finally some statistics are presented in Section \ref{SecStat},
 and concluding remarks in Section \ref{SecConclusion}.

\begin{table*}
\caption{The complete list of \xmm-LSS pointings in chronological order of observation.
\newline
Column (1) in each group of 4 is our own internal \textit{field name}
(the letter G refers to the Li\`ege/Milan/Saclay Guaranteed Time, the letter B to 
Guest Observer time, and the letter S to the SXDS; the suffix a,b,c indicates
repetition of a pointing because of insufficient exposure after high background
filtering). 
\newline
Fields flagged \textit{bad} in column (3) have usually been repeated
except for B17c, B45b, B47b, B68b which are the latest and best, though nominally
bad, and are necessary in order to avoid holes in the covered area.
\newline
Column (2) is the ESA ObsId identifier which can be used to look-up to the
pointing in the \xmmn~ log and archive.
\newline
The exposure (in ks) indicated in column (4) is 
the weighted mean of MOS1, MOS2 and pn nominal exposures.
These exposure times refer to \xlssd. For \xlssII~ all exposures longer than 10 ks
have been curtailed to such a length at event file generation time.
}
 \label{TabPointing} 
\begin{tabular}{lllr  lllr  lllr}
\hline (1) & (2) & (3) & (4) &
       (1) & (2) & (3) & (4) &
       (1) & (2) & (3) & (4) \\
\hline\hline
S01~ & 011237~0101 & $\dagger$ & 80.2$^{\dagger}$ & B20~ & 003798~2001 &     & 14.9 & B49~ & 040496~6301 &     & 10.5 \\
S02~ & 011237~0301 &     & 37.1 & B21~ & 003798~2101 &     & 12.6 & B51~ & 040496~6501 &     &  8.9 \\
S03~ & 011237~0401 &     & 15.7 & B26~ & 003798~2601 &     & 11.3 & B52~ & 040496~6601 &     & 12.5 \\
S04~ & 011237~1701 &     & 45.3 & B27~ & 003798~2701 &     & 13.8 & B54~ & 040496~6801 &     & 13.6 \\
G17~ & 011111~0301 &     & 20.6 & B17a & 003798~1701 & bad &  3.3 & B55a & 040496~6901 & bad &  6.6 \\
G18~ & 011111~0401 &     & 23.9 & B18~ & 003798~1801 &     & 13.7 & B56~ & 040496~7001 &     & 13.5 \\
G13~ & 010952~0501 &     & 21.4 & B19~ & 003798~1901 &     & 10.9 & B57~ & 040496~7101 &     & 10.5 \\
G19~ & 011111~0501 &     & 20.5 & G03~ & 011268~0301 &     & 20.7 & B59~ & 040496~7301 &     & 10.7 \\
G15~ & 011111~0101 &     & 16.9 & B25~ & 003798~2501 &     &  8.6 & B60~ & 040496~7401 &     & 13.5 \\
G16a & 011111~0201 & bad &  3.7 & B24~ & 003798~2401 &     & 14.0 & B61a & 040496~7501 & bad &  6.4 \\
G16b & 011111~0701 &     &  9.1 & B23~ & 003798~2301 &     &  7.9 & B62~ & 040496~7601 &     & 10.4 \\
B01~ & 003798~0101 &     & 11.6 & B22a & 003798~2201 & bad &  5.1 & B63~ & 040496~7701 &     & 12.6 \\
B06~ & 003798~0601 &     & 10.4 & B28~ & 014711~0101 &     & 10.4 & B64~ & 040496~7801 &     & 13.6 \\
B02~ & 003798~0201 &     & 10.5 & B29~ & 014711~0201 &     &  9.5 & B65~ & 040496~7901 &     & 13.5 \\
B07~ & 003798~0701 &     &  9.4 & B30~ & 014711~1301 &     & 11.4 & B66~ & 040496~8001 &     & 13.5 \\
B03~ & 003798~0301 &     & 10.7 & B31~ & 014711~1401 &     & 10.0 & B67a & 040496~8101 & bad &  4.8 \\
B05~ & 003798~0501 &     & 13.4 & B32a & 014711~1501 & bad &  1.6 & B68a & 040496~8201 & bad &  2.0 \\
B04a & 003798~0401 & bad &  5.9 & B42a & 040496~5601 & bad & 10.6 & B69~ & 040496~8301 &     &  8.7 \\
B09~ & 003798~0901 &     & 11.4 & B58a & 040496~7201 & bad &  6.0 & B70a & 040496~8401 & bad &  4.5 \\
G01~ & 011268~0101 &     & 24.2 & B44a & 040496~5801 & bad &  5.4 & B72~ & 040496~8601 &     & 11.8 \\
G04~ & 010952~0101 &     & 23.3 & B53~ & 040496~6701 &     &  9.4 & B71~ & 040496~8501 &     & 10.8 \\
G10~ & 010952~0201 &     & 22.0 & B48~ & 040496~6201 &     &  9.4 & B58b & 055391~1401 &     & 23.5 \\
G07~ & 011268~1001 &     & 22.2 & B04b & 040496~0101 & bad &  8.9 & B61b & 055391~1601 &     & 12.2 \\
G09~ & 010952~0601 &     & 19.5 & B13b & 040496~0201 & bad &  4.8 & B70b & 055391~1901 &     & 10.6 \\
G14~ & 011268~0801 &     & 11.0 & B17b & 040496~0301 & bad &  6.0 & B44b & 055391~0901 &     & 23.2 \\
G12a & 010952~0401 & bad &  1.7 & B32b & 040496~0401 &     & 10.5 & B45b & 055391~1001 & bad &  8.1 \\
G11~ & 010952~0301 &     & 19.3 & G12b & 040496~0501 &     &  9.5 & B46b & 055391~1101 &     & 21.1 \\
G05~ & 011268~0401 &     & 21.7 & B22b & 040496~0601 &     &  8.8 & B04c & 055391~0101 &     & 10.5 \\
B08~ & 003798~0801 &     &  8.7 & B33~ & 040496~4701 &     &  9.3 & B13c & 055391~0201 &     & 10.6 \\
G02~ & 011268~0201 &     &  8.9 & B34~ & 040496~4801 &     &  9.2 & B17c & 055391~0301 & bad &  7.8 \\
B10~ & 003798~1001 &     & 10.9 & B35a & 040496~4901 & bad &  5.1 & B35b & 055391~0401 &     & 10.0 \\
B11~ & 003798~1101 &     & 10.0 & B36a & 040496~5001 & bad &  8.1 & B47b & 055391~1201 & bad &  6.3 \\
B12~ & 003798~1201 &     &  9.4 & B37a & 040496~5101 & bad &  8.0 & B36b & 055391~0501 &     & 10.6 \\
B13a & 003798~1301 & bad &  4.4 & B38~ & 040496~5201 &     & 10.6 & B50b & 055391~1301 &     &  9.7 \\
B14~ & 003798~1401 &     &  9.3 & B39~ & 040496~5301 &     & 10.4 & B41b & 055391~0701 &     & 11.2 \\
G08~ & 011268~0501 &     & 19.0 & B40~ & 040496~5401 &     & 14.2 & B42b & 055391~0801 &     & 12.5 \\
B15~ & 003798~1501 &     & 10.3 & B41a & 040496~5501 & bad &  4.4 & B37b & 055391~0601 &     & 14.3 \\
B16~ & 003798~1601 &     & 10.2 & B43~ & 040496~5701 &     & 13.4 & B55b & 055391~1501 &     & 12.6 \\
G06~ & 011268~1301 &     & 13.4 & B45a & 040496~5901 & bad &  6.8 & B67b & 055391~1701 &     & 12.4 \\
S06~ & 011237~0701 &     & 46.9 & B46a & 040496~6001 & bad &  7.0 & B68b & 055391~1801 & bad &  4.9 \\
S07~ & 011237~0801 &     & 35.8 & B47a & 040496~6101 & bad &  5.9 \\
S05~ & 011237~0601 &     & 32.7 & B50a & 040496~6401 & bad &  4.8 \\
\hline
\end{tabular}

\raggedright
$\dagger$ 
For field S01 the full exposure is much longer than the typical \xmm-LSS exposure
and for this reason the relevant data are fictitiously
flagged bad in \xlssd, while those
deriving from an analysis curtailed at 40 ks are used instead, 
with a field name of  S01\_40 in column (1) and an exposure of  40.0 ks
in column (4).
\end{table*}

\begin{figure*}
 \includegraphics[width=168mm]{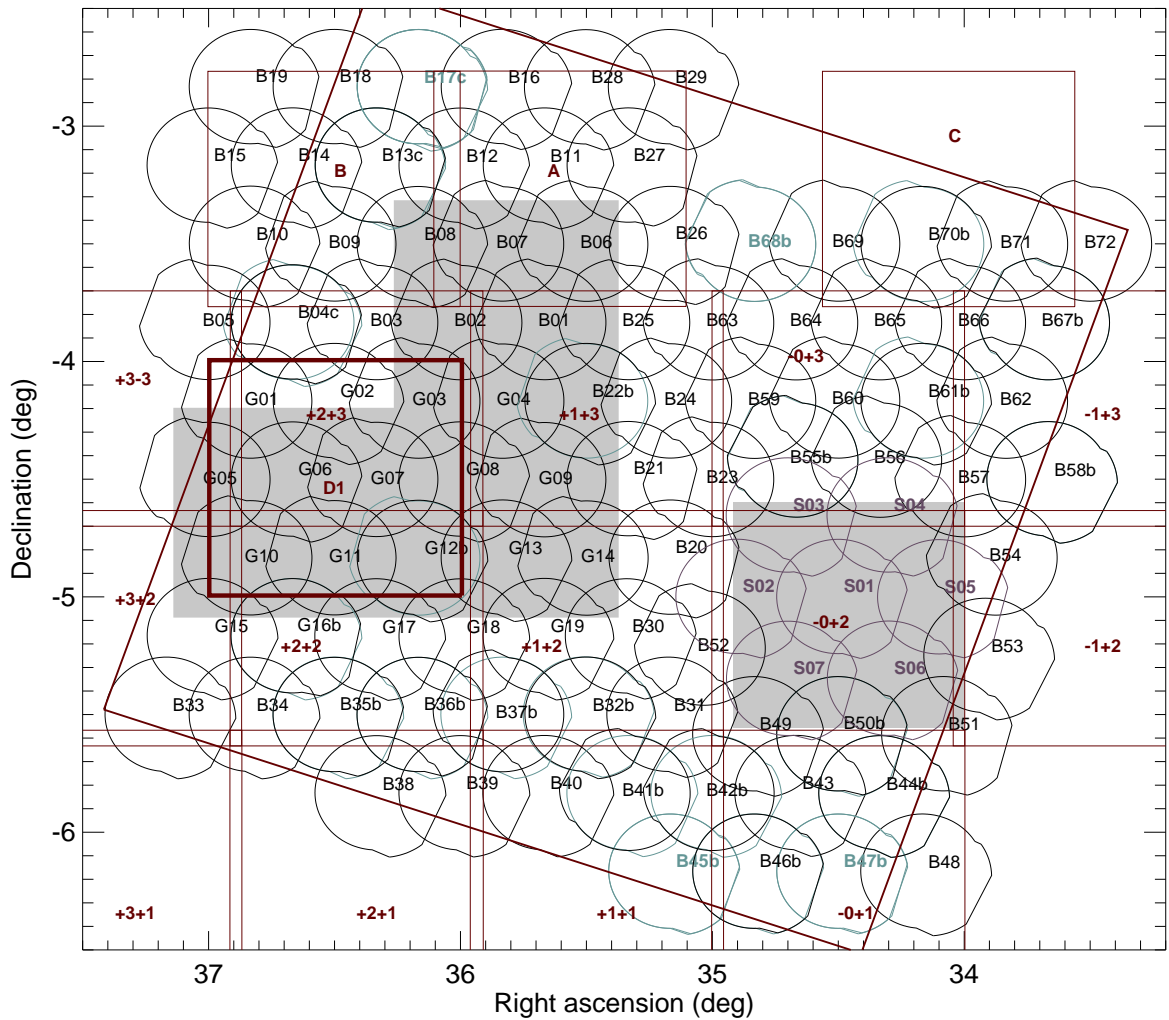} 
 \caption{Layout of the \xmm-LSS pointings and coverage in other wavebands.
 The position of the \xmm~ FOV footprints of good pointings are plotted and
 labelled with their field name (in black). SXDS pointings are plotted and labelled
 in gray (pink-gray in the web version). Bad pointings (later repeated by a good
 one) are plotted in light gray (blue-gray in the web version), without labels
 (except for the 4 cases where even the last re-observation is nominally bad).
 The total geometrical area is estimated to be 11.1 \dd.
        The dark gray (maroon in the web version) squares indicate the various tiles of the CFHTLS W1 survey
        (labelled with their short $\pm x \pm y$ name; see
        \url{http://terapix.iap.fr/cplt/oldSite/Descart/cfhtls/cfhtlswidemosaictargetW1.html}),
        of our own supplementary pointings (labelled ABC) and
        the CFHTLS D1 field (thick). The large tilted square is the area covered
        by the SWIRE survey. The shaded areas are those covered by the UKIDSS surveys
        (DXS, left; UDS, right) in release DR5.
        The entire area is covered by the \textit{GALEX} AIS and DIS surveys.
 }
 \label{FigPointing}
\end{figure*}

\section{Catalogue layout and content}
\label{SecNEWCat}

\subsection{List of available pointings}
\label{SecPointings}

The entire \xmm-LSS survey 
consists of 91 positions on the sky arranged with
a regular spacing. Some of the pointings were however repeated once or twice, because
the first observations were flagged bad due to a too high background or insufficient
clean exposure time.  A total of 117 pointings were executed during the
Guaranteed Time,  AO-1, AO-2, AO-5 and AO-7 periods. In addition 7 pointings of
the independent Subaru \xmmn~ Deep Survey
(SXDS; \citealt{2008ApJS..179..124U}), with a somewhat different
spacing but fully surrounded by our own pointings, were retrieved from the archives and 
reanalysed by us with our pipeline.
The complete list of all 124 pointings is given in Table~\ref{TabPointing}, while
the layout on the sky is plotted in Fig.~\ref{FigPointing}.

\subsection{The X-ray source lists}
\label{SecXlist}

In this paper we present two variants of the X-ray catalogue, each including 
source lists for two bands, 0.5-2 and 2-10 keV, named B and CD respectively.
The \textit{deep} catalogue (\xlssd) is obtained from the
processing of event files for the entire exposure of each pointing (with the
exception of pointing S01, whose duration is much longer than all other pointings, and
which has been processed also as a "chunk" of 40 ks).
The \textit{standard} catalogue (\xlssII) instead uses a uniform exposure of 10 ks 
for all pointings longer than that.
Both catalogues share an identical processing and the same layout.
The list of database columns are reported in Tables \ref{TabBand} and \ref{TabMerge}
in Appendix~\ref{SecAPPTables}.

\subsection{The multiwavelength catalogues}
\label{SecMultilambdaFiller}

We provide also multiwavelength catalogues, named in the database
{\tt 2XLSSOPT} and {\tt 2XLSSOPTd} (see
list of database columns in Table \ref{TabMulti}
in Appendix~\ref{SecAPPTables}), generated correlating
the X-ray source list with optical, NIR, IR and UV catalogues as described in
Section~\ref{SecOpt}.

\subsection{Summary of online availability}
\label{SecOnline}

\subsubsection{The database tables}
\label{SecOnline1}

\begin{table}
\caption{The database tables of the current release}
\label{TabTables}
\begin{tabular}{ll}
 \hline
 Data sets & Tables \\
 \hline\hline
 10 ks catalogues \\
 \hline
 Merged  catalogue (all parameters): & {\tt 2XLSS} \\
 Single-band  catalogues: & {\tt 2XLSSB, 2XLSSCD } \\
 Multiwavelength catalogue: & {\tt 2XLSSOPT} \\
 \hline
 deep catalogues \\
 \hline
 Merged  catalogue (all parameters): & {\tt 2XLSSd} \\
 Single-band  catalogues: & {\tt 2XLSSBd, 2XLSSCDd } \\
 Multiwavelength catalogue: & {\tt 2XLSSOPTd} \\
 \hline
 XMDS catalogues \\
 \hline
 Multiband X-ray catalogue: & {\tt XMDS}    \\
 Multiwavelength catalogue: & {\tt XMDSOPT} \\
 \hline
\end{tabular}
\end{table}

The database site at IASF Milano described in Paper I was relocated since August 2007 to
the new site  \url{http://cosmosdb.iasf-milano.inaf.it/XMM-LSS/}, and converted to
the \textit{DART} interface \citep{2008ASPC..394..397P}
developed by us and used at IASF to support several other projects.
While the underlying MySQL database structure is virtually unchanged since
the one described in \cite{2005A&A...439..413C}, the user interface has been improved
and in particular now requires public users to register with an individual username (see instructions
reachable from the home page).

In addition to the material described in Paper I (which will continue to remain
available), the database tables listed in Table~\ref{TabTables} 
(plus the data products described in Section \ref{SecDP}) will be available 
in our database allowing fully interactive selection.
Refer to Appendix~\ref{SecAPPTables}
for the subset available in electronic form also at CDS.

Single-band tables are provided separately for the B [0.5-2] keV and CD [2-10] keV bands.
They contain a selection of parameters generated by \xamin, like
both sets of values computed for the point-like and extended source fit.
Position errors and fluxes are derived a posteriori, and
computed as per Section~\ref{SecFlux}.
Only sources above a detection likelihood of 15 are made available in the single-band tables.
Redundant sources detected in overlapping regions of different pointings are
removed as explained in Section~\ref{SecOverlap}; 

The B-CD band merged catalogue is obtained matching single band detections within
a correlation radius of 10\arcsec~ (see Section~\ref{SecMerging}), and includes only
the parameters for the classification (point-like or
extended) relative to the \textit{best band}. Data in the other band are made
available even if they have a detection likelihood below 15.

\subsubsection{Associated data products}     
\label{SecDP}

Data products are files associated with a given database entry. We distinguish the
case of \textit{per-pointing} and \textit{per-object} data products. When a database
query returns a number of X-ray sources, each of them may point to an individual
data product, or to one common to the pointing where the source was detected.
The database interface allows the user to retrieve individual data products, or to build
on the fly a \texttt{.tar.gz} file containing all the products related to the query.

\paragraph{X-ray images}         
\label{SecXDP}
~\newline

The following X-ray data products are available \textit{per-pointing}
for the deep catalogue only:

\begin{enumerate}
\item The B and CD  band  photon  images (one mosaic cumulative for the 3 detectors,
      after the event filtering)
\item The B and CD wavelet images derived from the above
\item Separate exposure maps for the 3 detectors and 2 bands.
\item ds9 contours (log-spacing based on B band wavelet images)
\end{enumerate}

All images
have a pixel size of 2.5\arcsec. Note that the World Coordinate
System (WCS) of the X-ray images is the one generated by the {\sc SAS}, therefore it does not take into account the astrometric
correction described in Section~\ref{SecAstro}.
Consequently when overlaying X-ray source
positions exactly on  the X-ray images, one should use the
coordinates labelled as ``raw" in Table \ref{TabBand}, although
this  does not make much difference for most of the sources, given
the pixel size.

\paragraph{Multiwavelength data}   
\label{SecODP}
~\newline

The following thumbnail images are available \textit{per X-ray source} in
association with the band merged and multiwavelength catalogues (deep version only).
The FITS thumbnail images have proper WCS which allows direct overlaying of X-ray
astrometrically corrected positions as well as counterpart positions.

\begin{enumerate}
\item FITS CFHT images ($40\arcsec \times 40\arcsec$)  in the $i'$ band for sources
      covered by the CFHTLS W1 or D1 fields and/or in the
      $g'$ band when covered by our own ABC fields,
      obtained via the CADC cutout service
      \footnote{\url{http://www.cadc-ccda.hia-iha.nrc-cnrc.gc.ca/}}.
      A PNG version is also available as for Paper I.
\item FITS SWIRE images in the 4 IRAC bands ($30\arcsec \times 30\arcsec$)
      and in the 3 MIPS bands ($60\arcsec \times 60\arcsec$),
      obtained via the IPAC Gator cutout service
      \footnote{\url{http://irsa.ipac.caltech.edu}}
\end{enumerate}

\section{X-ray data processing}
\label{SecNEWProc}

The original \xamin~ pipeline used in Paper I was described in detail in
\cite{2006MNRAS.372..578P}.
While referring to such papers for detail, we summarize here the main
processing steps.

Standard {\sc SAS} tasks are used to generate event lists. They are filtered
for solar soft proton flares and used to produce images for the three {\it EPIC}
detectors, which are then co-added in each energy band. 
Such per-band images are filtered in wavelet space, and scanned by a source detection algorithm
based on {\sc SExtractor} \citep{1996A&AS..117..393B}
to obtain a primary source list. Source characterization is
then performed with \xamin, a maximum likelihood profile fitting procedure, designed 
for the \xmm-LSS survey, optimized for extended X-ray sources and associated 
signal to noise regimes.
\xamin~ performs parallel fits with two classes of surface-brightness
models, a point-like one and an extended
($\beta-$profile) one and outputs the main parameter for both models in a FITS table
per pointing and per band. Further processing (as described in Sections
\ref{SecAstro}, \ref{SecMerging} and \ref{SecOverlap}) is performed contextually or
after ingestion of \xamin~ output into our database.

\subsection{The revised pipeline}
\label{SecXamin}

\begin{figure}
 \includegraphics[width=8.5cm]{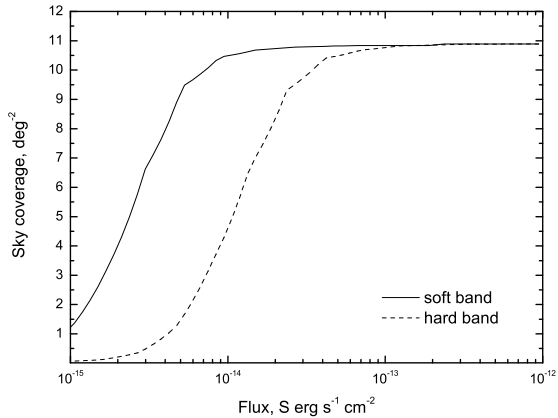} 
 \caption{Effective sky coverage for the entire \xmm-LSS area in the soft (B, [0.5-2] keV)
  and hard (CD, [2-10] keV) bands.
  This figure, reproduced from Fig. 9 of Elyiv et al. (2012), 
  updates and supersedes Fig. 2 of Paper I.
  The computation derives from simulations of realistic \xmm-LSS observations
  processed by \xamin.
 }
 \label{FigElyiv9}
\end{figure}

For this release, we used the latest version \textit{pro tempore} (3.2) of
\xamin~, which has been improved while translating it from IDL to Python (open source).
All pointings, including those reported in the \xlss~ catalogue in Paper I,
have been (re)processed afresh with this latest version.

 The \xamin~ pipeline output parameters for version 3.2 are the same listed
 in Table 2 of Paper I
 (and flagged in column (X) in Tables \ref{TabBand} and \ref{TabMerge}).

 The event file generation (and the subsequent pipeline) was applied independently 
 to the full exposure of each pointing, as well as to
 10 ks curtailed chunks (from the beginning of the exposure).
The latest \xmm~ calibrations available \textit{pro tempore} were applied.

One of the differences between the old and new pipeline is the correction
of an offset of 0.5 pixel (where our pixel size is 2.5\arcsec) in  XY image
positions. For this reason all X-ray source positions and catalogue names
(see Section \ref{SecNaming}) have changed.

We checked
that the new pipeline version provides results consistent with the previous IDL
version by performing detailed tests on simulated and real \xmm~ pointings; then
we proceeded to a direct comparison, which shows a
(good) agreement between the old and new pipeline as
reported in Appendix~\ref{SecXstatNewOld}. 

This is a summary list of all differences in catalogue generation
with respect to Paper I.

\begin{enumerate}
 \item more input data (5 to 11 \dd)
 \item used latest {\sc SAS} version and calibrations
 \item used \xamin~ version 3.2
 \item in particular half-pixel offset cured (see above)
 \item astrometry using CFHTLS T004 (see Section \ref{SecAstro})
 \item band merging at 10\arcsec~ (see Section \ref{SecMerging})
 \item overlap removal at 10\arcsec~ (see Section \ref{SecOverlap})
 \item web site relocated (see Section \ref{SecOnline})
 \item more multi$-\lambda$ bands (see Section \ref{SecOpt})
\end{enumerate}

In addition to the {\sc Xamin} output, a number of parameters are calculated
a posteriori in order to facilitate the interpretation
of the data set. Since in its present state, {\sc Xamin} does not
perform error calculations, mean statistical  errors were
estimated by means of extensive simulations, as explained in Paper I and
\citet{2006MNRAS.372..578P};
we note that  only the first 2 digits are to be
considered significant for the count rate and for the core radius
as well as for the derived quantities.

Analogously to Paper I, only sources with an off-axis angle $<13\arcmin$ are
processed by \xamin.
The catalogues include all the extended sources classified
in the customary C1 and C2 classes (see 
Section~\ref{SecC1C2}) 
plus all point-like
sources with a point source detection likelihood ($LH$) greater than 15 (so-called 
{\it non-spurious}).
The resulting sky coverage is shown in Fig.~\ref{FigElyiv9}.

   \begin{figure*}
   \centering
   \includegraphics[width=8.5cm]{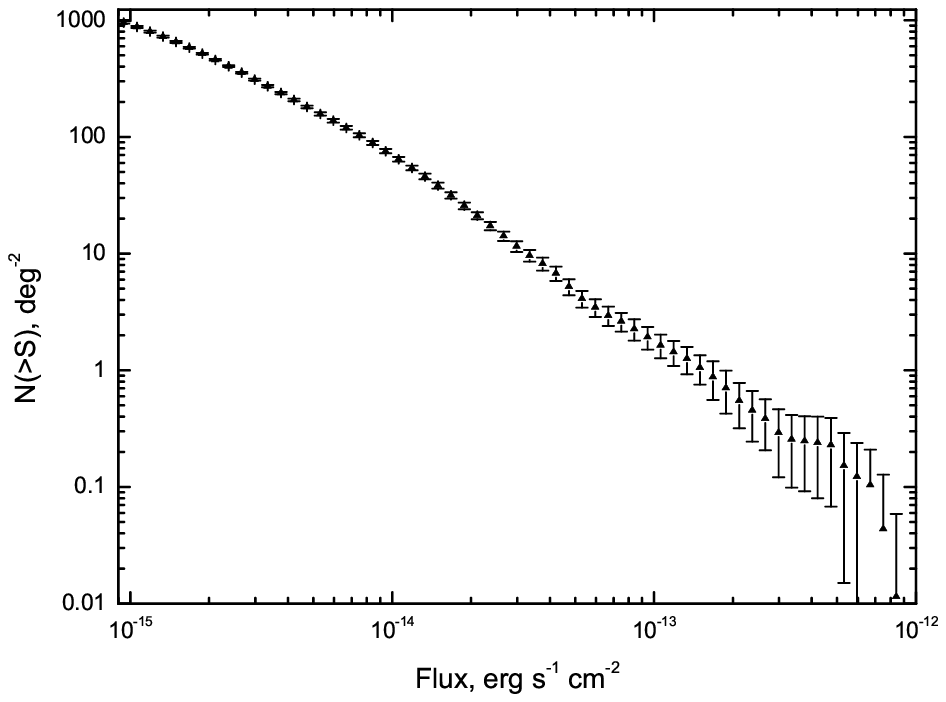} 
   \includegraphics[width=8.5cm]{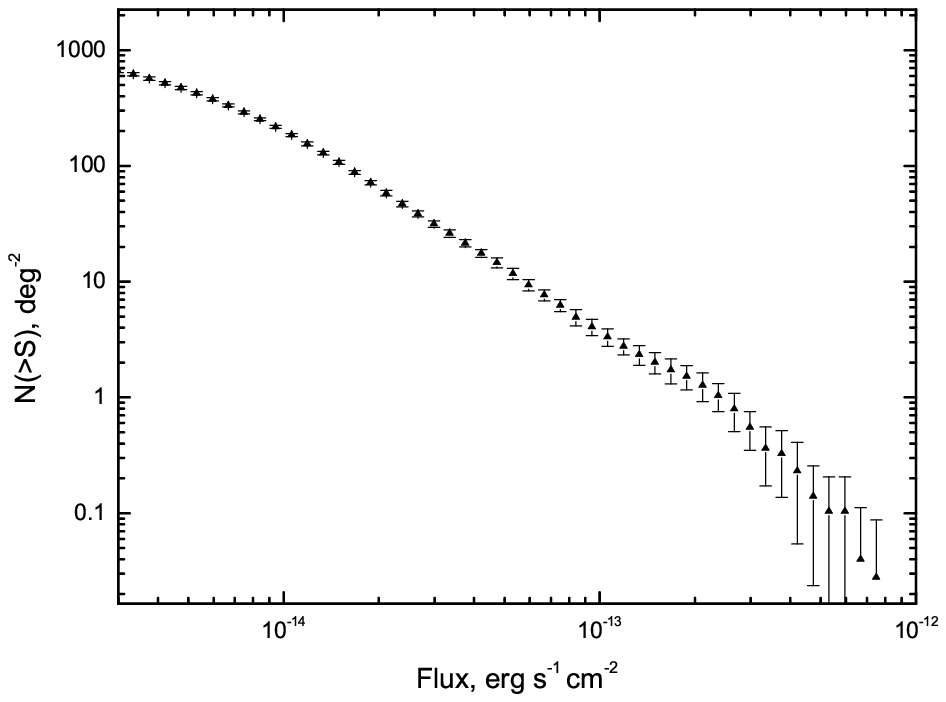} 
   \caption{The logN-logS distributions for the soft (left-hand panel) and hard
    (right-hand panel) for the entire \xmm-LSS area. These figures are reproduced
    in a simplified form from
    Figs. 13 and 14 of Elyiv et al. (2012) 
             }
   \label{FigElyivLNLS}
    \end{figure*}

\begin{figure*}
\centerline{
 \includegraphics[width=6cm]{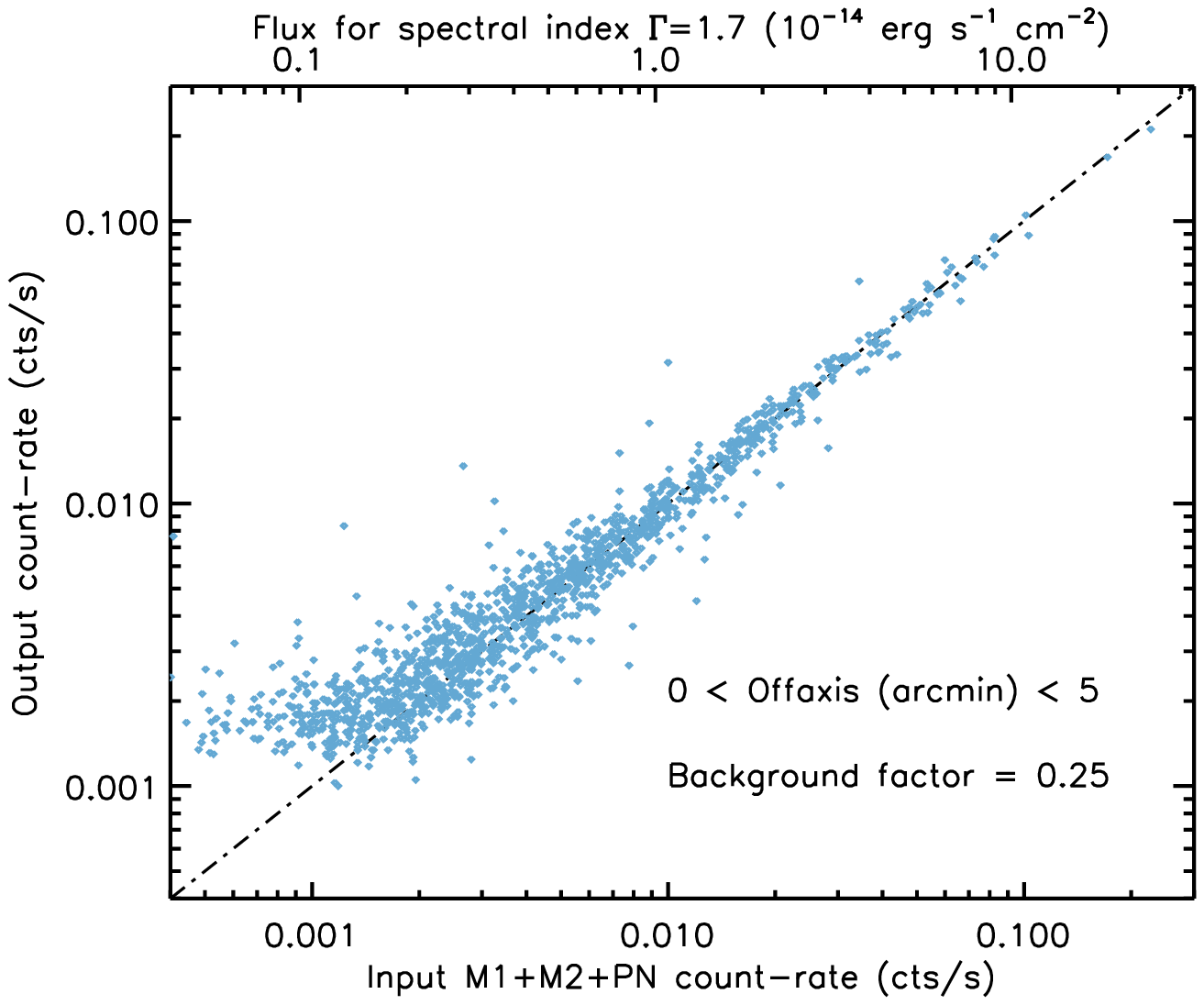}  
 \includegraphics[width=6cm]{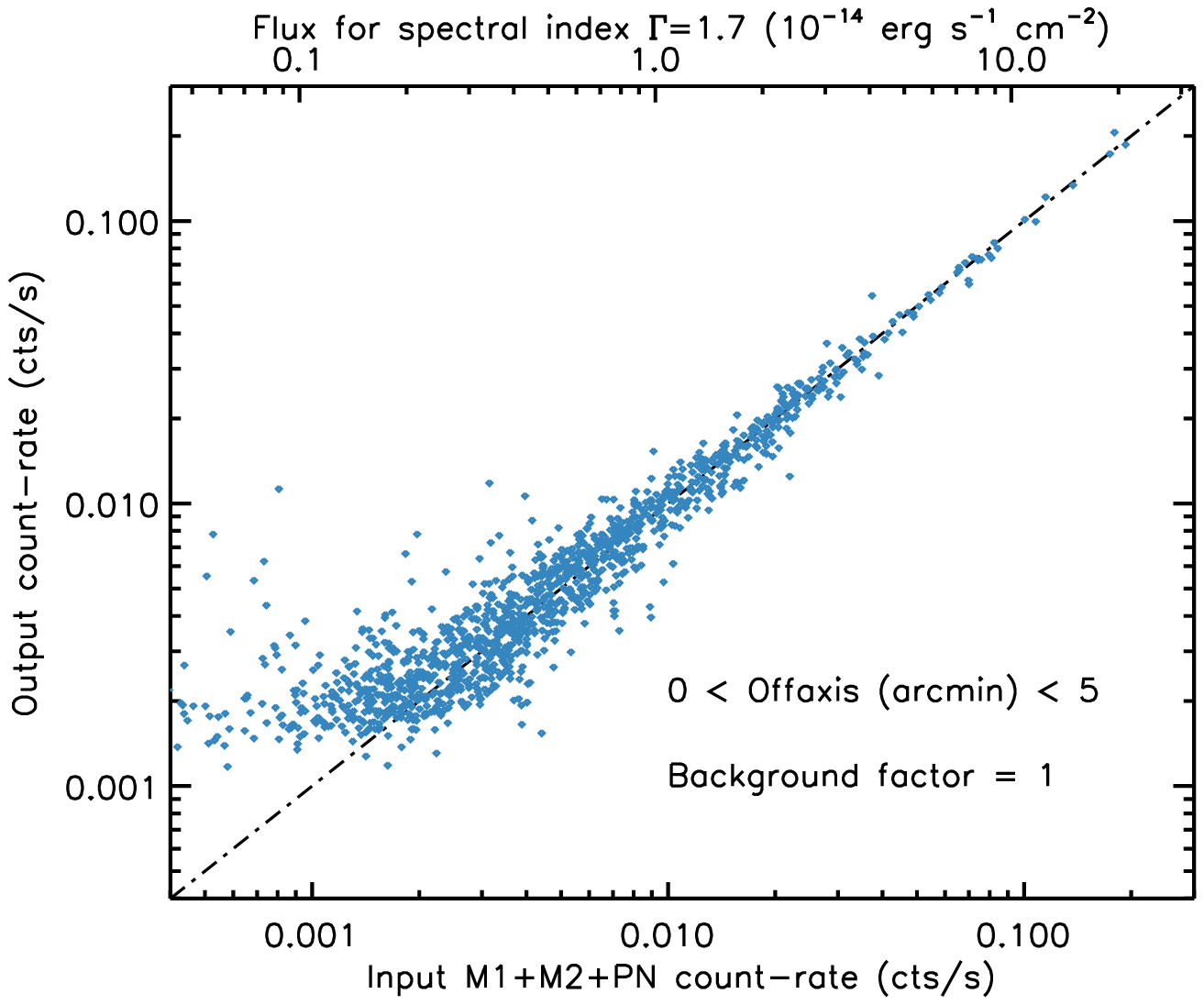}  
  \includegraphics[width=6cm]{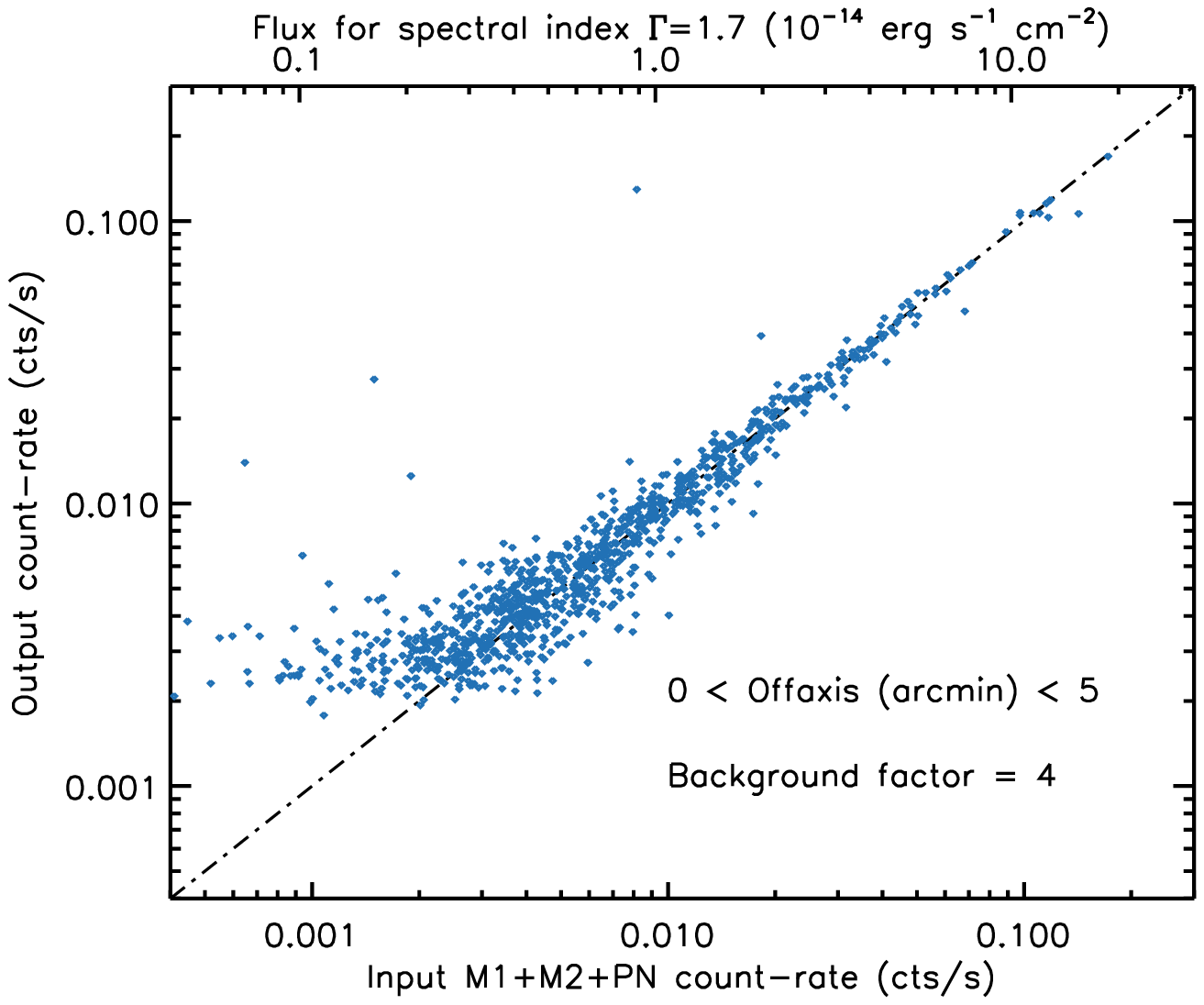} 
 }
\centerline{
 \includegraphics[width=6cm]{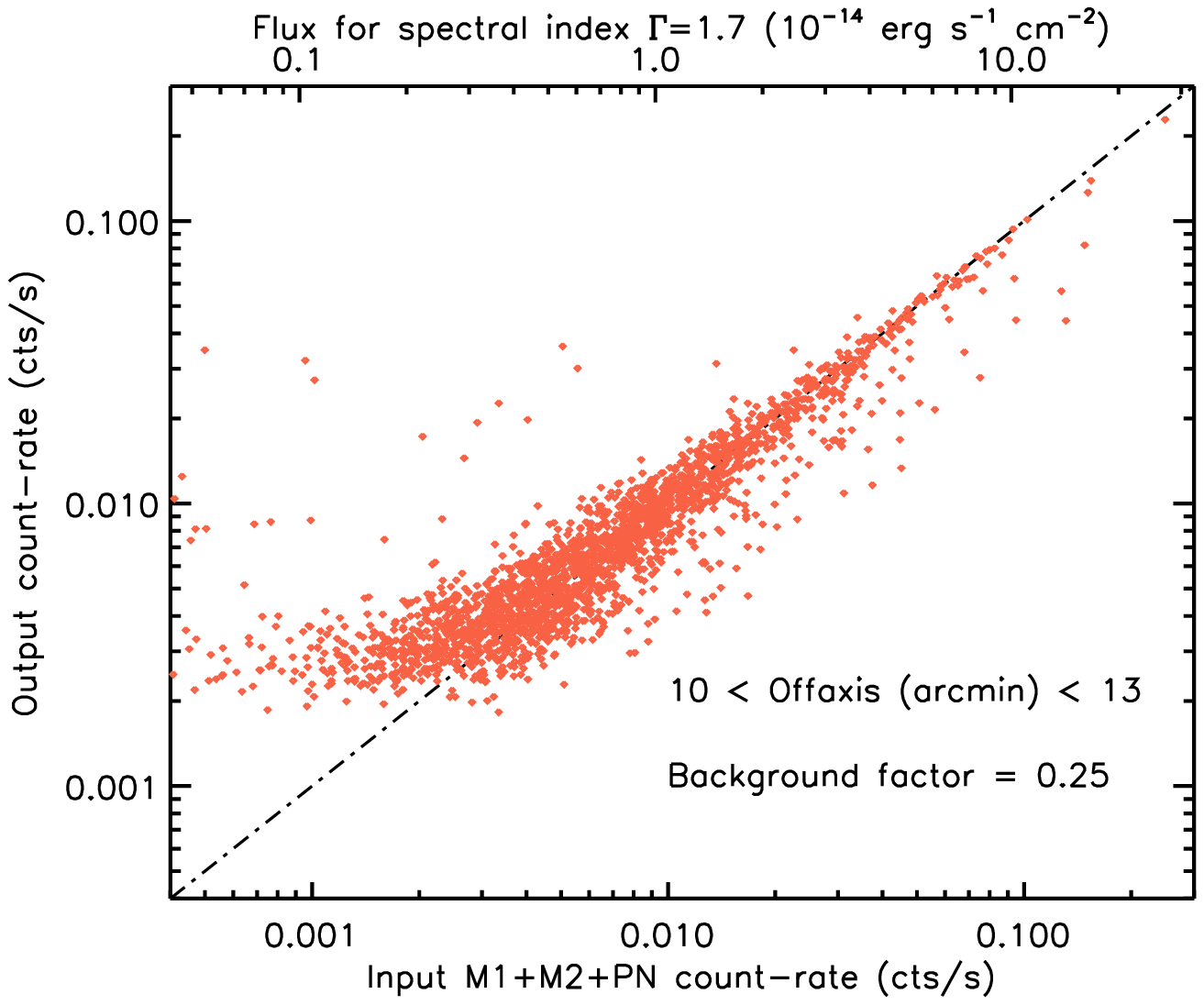}  
 \includegraphics[width=6cm]{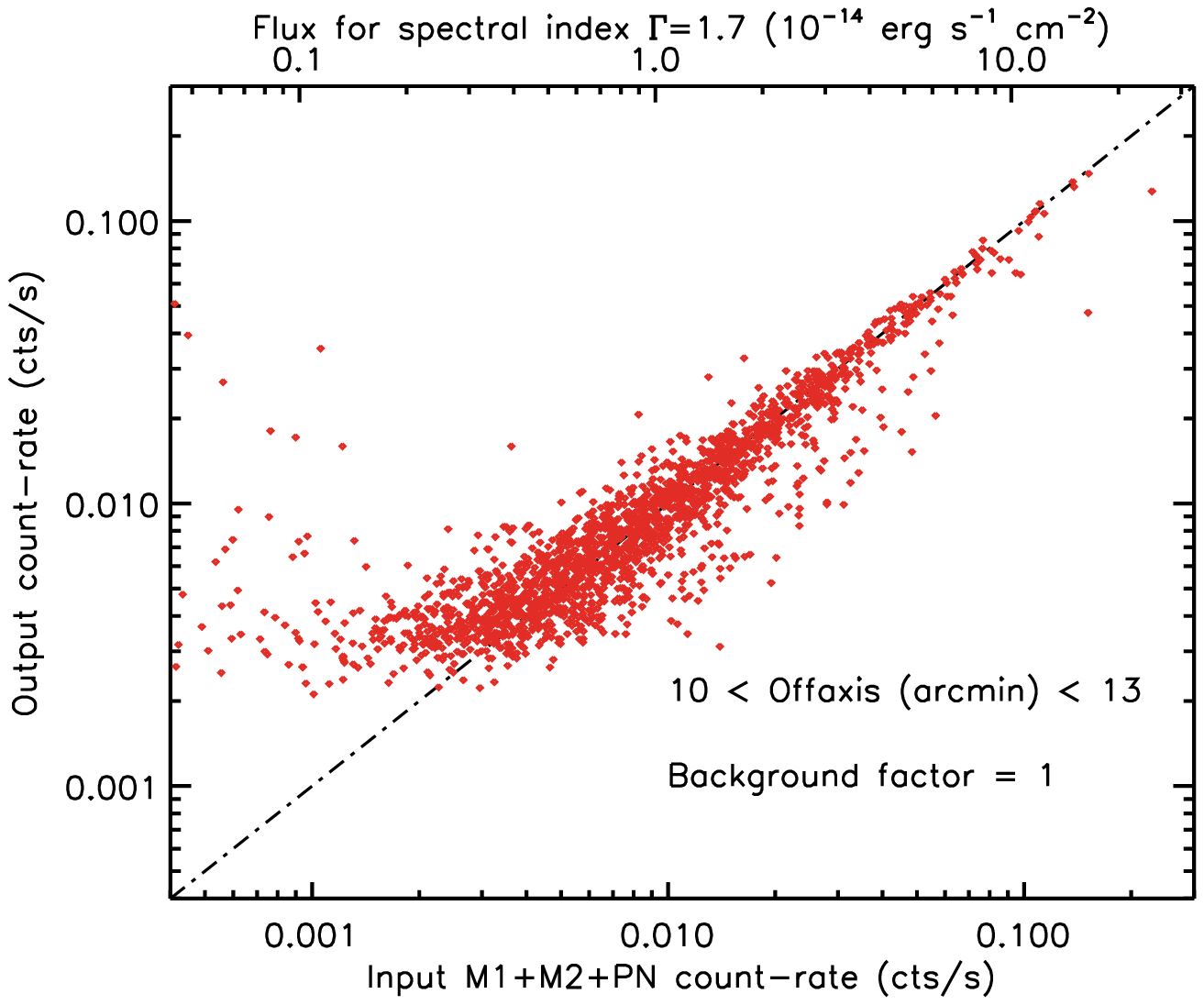}  
  \includegraphics[width=6cm]{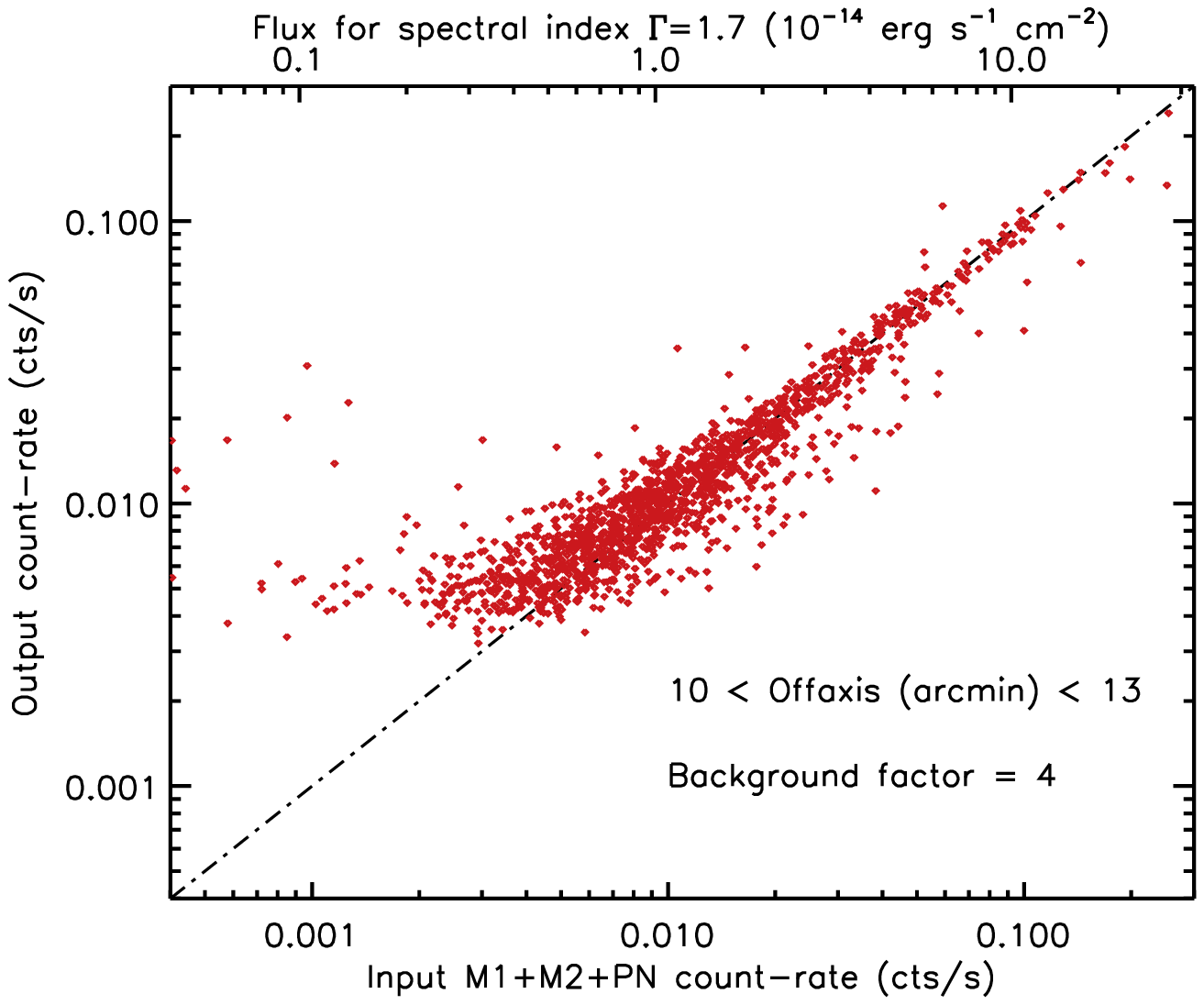} 
 }
\caption{Photometric accuracy for three background values (see text) 
 and two off-axis angle ranges (top:
 0-5\arcmin~ and bottom: 10-13\arcmin;
 plots for the intermediate} range are rather similar).
``Count-rate'' is the measured MOS1+MOS2+pn rate, normalised to the on-axis value.
\label{FigClerc}
\end{figure*}

\begin{table}
\caption{
The  Energy Conversion Factors for the individual EPIC
cameras and energy bands, in units of $10^{-12}$ \flux\ for
a rate of one count/s.  A photon-index power-law of 1.7 and a mean
$N_{H}$ value of $2.6~ 10^{20}$ cm$^{-2}$ are assumed. The two
MOS cameras are assumed to be identical. 
\label{TabCF}}
\begin{tabular}{lll}
   \hline
   Detector & B band & CD  band\\ \hline\hline
   MOS & 5.0 & 23\\
   pn & 1.5 & 7.9 \\
   \hline 
\end{tabular}
\end{table}

\subsubsection{Countrate and Flux}
\label{SecFlux}

As in Paper I, fluxes are not computed by \xamin~ but are inserted
in the catalogue as \textit{derived parameters}, i.e. a
single mean flux ([FLUX(MOS)+FLUX(pn)]/2) is computed from the count rates
using the customary conversion factors 
reported in Table~\ref{TabCF}, assuming the spectral model given in its caption.

The observed logN-logS distributions are presented in 
Fig.~\ref{FigElyivLNLS} reproduced in a simplified form from
\cite{2012A&A...537A.131E}.

Photometric accuracy plots based on simulations were presented in Fig.~3 of Paper I 
as a function of different off-axis angle ranges.
Further simulations were performed considering
different background levels 
as an additional parameter and are presented, supplementing Paper I,
in Fig.~\ref{FigClerc} (the "background factor" of 0.25, 1 and
4 refer to the nominal particle background defined in Table 1 of
\cite{2012A&A...537A.131E};
the latter paper, to which we refer for details, gives also
an alternate view in its Figs.~11 and 12).
We conclude that
the extremely weak dependency on background (if any) does not require to introduce
it into
the parametrization of photometric bias and accuracy as a function of count rate
and off-axis angle
published in Table 6 of Paper I.

\subsection{Positional accuracy and  astrometric corrections}
\label{SecAstro}

The \xamin~ pipeline does not provide directly error values,
since, for efficiency purposes, the likelihood surface is only searched
for its maximum.
Therefore, identically to what was done in Paper I,
the positional (\textit{statistical}) error on the (point-source) coordinates 
is estimated from Monte-Carlo simulations, and the values
indicated in the catalogue are computed from a look-up table  
of discrete values as a function of
count rate and off-axis angle ranges (as reported in 
Table~\ref{Tab8PapI}).
The distribution of the errors is shown in panel (b) of Fig.~\ref{FigDistHisto}.

Similarly to Paper I, in order to compensate for possible \textit{systematic} 
inaccuracies in the \xmm~ pointing positions, a global rigid astrometric correction
was estimated using the {\sc SAS} task {\sc eposcorr}
(with rotational offset search disabled). The correction offsets
were computed afresh for the full exposure case, and applied to both the \xlssd~ and \xlssII~
catalogues.

The input to {\sc eposcorr} were, for each pointing, an X-ray reference file with
all \textit{non-spurious} X-ray sources, while optical reference files were generated
taking all objects in the CFHTLS W1 fields within 6\arcsec~ from the (raw) source position,
brighter than $i'=25$ (or $r'=25$ for the ABC fields, see Section~\ref{SecOptInput}),
and having a chance probability (as defined in Section~\ref{SecOptProb}) $prob < 0.03$.
In case of more possible counterparts, the one with the smallest probability was taken.
Fields B68a and B68b (both bad) had no CFHT counterparts and were corrected using stars
in USNO A2.0. Field G12a (bad) had no counterparts at all and was not corrected.

The offsets computed by {\sc eposcorr}  were applied to all coordinate sets for
each source in the database. Astrometrically corrected positions were used  in the
subsequent operations: removal of the redundant sources, source naming and 
cross-identification with the catalogues in other wavebands.

In most cases the offsets are rather small and barely 
significant\footnote{tabulated online at \url{http://cosmos.iasf-milano.inaf.it/\~lssadmin/Website/LSS/List/.newastroreport.html}}
The range of the RA offset is $-3.7\arcsec < \Delta RA < 1.1\arcsec$ (with just 16
pointings with $|\Delta RA| > 2\arcsec$, 27 pointings with a significance of the offset
greater than $3\sigma$, of which 13 above $4\sigma$).
The range of the Declination offset is $-2.7\arcsec < \Delta Dec < 2.7\arcsec$ (with
just 4 pointings with $|\Delta Dec| > 2\arcsec$,  7 pointings with a significance greater
than $3\sigma$, of which 3 above $4\sigma$).

The quality of the positional accuracy can be estimated \textit{a posteriori}
from figures like Figs.~\ref{FigDistHisto} and \ref{FigAstroDist}.
For a final statistics see Section~\ref{SecOstat}.

\begin{table}
\caption{Positional accuracy (1$\sigma$ error on R.A. or Dec.)
for point sources derived from simulations of 10 ks pointings and
having a detection likelihood $> 15$.  Values are
given for the B and CD bands, as a function of the  summed
measured count-rate: $CR$ = MOS1+MOS2+pn.
This table is reproduced from Table~8 of Paper I, to which the reader is referred
for further details.}
\label{Tab8PapI}
 \centering
\begin{tabular}{l c c}
 \hline \hline
Band & B & CD \\
  Countrate (count/s) &  Error   ($''$) & Error  ($''$)\\
 \hline
 $0 <$off-axis$<5'$   & & \\
$0.001<CR<0.002$ & 2.0 & 2.0\\
 $0.002<CR<0.005$ & 1.7 & 1.7 \\
 $0.005<CR<0.01$ &1.3& 1.3 \\
 $CR>0.01$ & 0.8 & 0.8\\  \hline
 $5'<$off-axis$<10'$   & &\\
 $0.001<CR<0.002$ & 2.0 & 2.0\\
 $0.002<CR<0.005$ & 1.8 & 1.9\\
 $0.005<CR<0.01$ &1.5 & 1.5\\
 $CR>0.01$ & 1.0 & 1.0 \\ \hline
 $10'<$off-axis$<13'$   & \\
 $0.001<CR<0.002$  & - & -\\
 $0.002<CR<0.005$ & 1.9 & 2.0\\
 $0.005<CR<0.01$ & 1.6 & 1.7\\
 $CR>0.01 $& 1.2 &1.3 \\
 \hline
\end{tabular}
\end{table}

\subsection{Band merging}
\label{SecMerging}

The \xamin~ pipeline has been
optimized for the detection of clusters (which occurs preferentially in the soft band),
and its wavelet filtering component is inherently working on a single band
(see \cite{2006MNRAS.372..578P} and references therein), therefore
it is natural that energy bands are treated separately and the
merging is performed at the post-processing stage, namely
in the database ingestion stage.
Since, as in Paper I, we intend to provide
an X-ray band-merged catalogue along with the single-band ones, such a
merging procedure 
was defined in Paper I to cope with the case that
an X-ray source can be detected in one or two bands and, for
each band, can be independently fitted by the extended and point
source models with the coordinates free.
For each band, a source is classified as
extended (E) as described in Section~\ref{SecC1C2}, otherwise
it is classified as point-like (P). Then,  pointing by
pointing, we flag associations between the 2 bands within a search
radius of 10 \arcsec.  Note that we allow associations involving
spurious sources ($LH<15$) at most in one band. We keep the
information (rate, flux, etc.) about entries below this threshold
in the merged catalogue, since it could be more useful (e.g. for
upper limits) than no information at all, but we flag those cases
with ${\tt Bspurious}=1$ or ${\tt CDspurious}=1$. Finally, for each
soft-hard couple in the merged catalogue, we define the {\it best
band}, i.e. the band in which the detection likelihood of the
source is the highest and from which the coordinates are taken. 
The source flagging and classification (and the way fluxes appear
in the database) is identical to the one described in Tab. 9 of Paper I.

\textit{The change with respect to Paper I is the increase of the search
radius from 6 to 10\arcsec.}
In fact an examination of the \xlss~ catalogue showed an excess of couples of
soft-only and hard-only sources usually detected in the same pointing
with a distance marginally above 6\arcsec. They could be interpreted as
"potential missed mergers" since they might have escaped band-merging because
of the distance. Or, if they were in different pointings, they could be "potential
missed overlaps". 
We performed a thorough analysis 
of sources closer than 30\arcsec.
95\% of the detections in the same field before band merging, closer than 10\arcsec, meet the definition of missed mergers,
while only 25\% of those farther than 10\arcsec~ do.

The \textit{starting point} is represented by the individual band tables.
After the initial merging procedure for \xlssd~
one can directly compare the 10\arcsec~ and 6\arcsec~ merging as shown in
Table~\ref{TabStatMerge}, where
(a) \textit{"preserved"} means they are either
unmerged (single band detection) or merged in the same way, and identical in all respects;
(b) \textit{"upgraded"} means they would have been
considered at 6\arcsec~ as detections in a single band, and are merged into one at 10\arcsec;
(c) lost means single-band detections at 6\arcsec no longer considered.

As already described in Paper I, there is a limited number of cases where the band
merging is \textit{primarily} ambiguous, and a source in a band happens to be associated with two
different objects in the other band (i.e. gives rise to a couple of entries in the merged table). 
The implication on source naming is discussed below in Section~\ref{SecNaming}. 
In a further step of the band merging procedure we also considered \textit{secondary}
ambiguous cases based on the inter-band distance (database column ${\tt Xmaxdist}$)
between the positions found by {\sc Xamin} in the two energy bands: if in a couple both
${\tt Xmaxdist}<6\arcsec$ (i.e. they would have been ambiguous also with the old
merging), or both ${\tt Xmaxdist}>6\arcsec$ (irremediably ambiguous), both 
merged entries are maintained; when one ${\tt Xmaxdist}$ is below 6\arcsec~ and the
other above, the latter entry is \textit{divorced}. The lower-distance element
remains a merged two-band detection, while the other is reset to an only-hard or
only-soft source. 

\begin{table}
\caption{Statistics of the band merging procedure}
\label{TabStatMerge}
\begin{center}
\begin{tabular}{lrr}
\hline
 Number of sources for condition             & total   & non-spurious \\
\hline\hline
in input soft-band table                     & 10348   &         7339 \\
in input hard-band table                     &  7124   &         3235 \\
after initial merging                        & 14216   \\
~of which preserved w.r.t. 6\arcsec~merging  & 13713   &         7824 \\
~of which upgraded                           &   505   &          457 \\
~lost (no longer considered)                 &   492   \\
${\tt Xmaxdist} < 2\arcsec$                  &   33\%  &         37\% \\
${\tt Xmaxdist} < 4\arcsec$                  &   69\%  &         75\% \\
\hline
after ``divorce" procedure \\
~preserved                                   & 13724   \\
~upgraded                                    &   492   \\
\hline
\end{tabular}
\end{center}
\end{table}

Note that not all sources in the produced merged table will go
into the catalogue: those which are spurious in both bands will not go, as well as those
removed as redundant according to the procedure in Section~\ref{SecOverlap}.
The total number of ambiguous cases in the final catalogue is really
small: 20 couples and 5 singles (over 6721) for \xlssd~ and 15 couples and 8 singles
(over 5572) for \xlssII.

\begin{figure}
   \includegraphics[width=8.4cm]{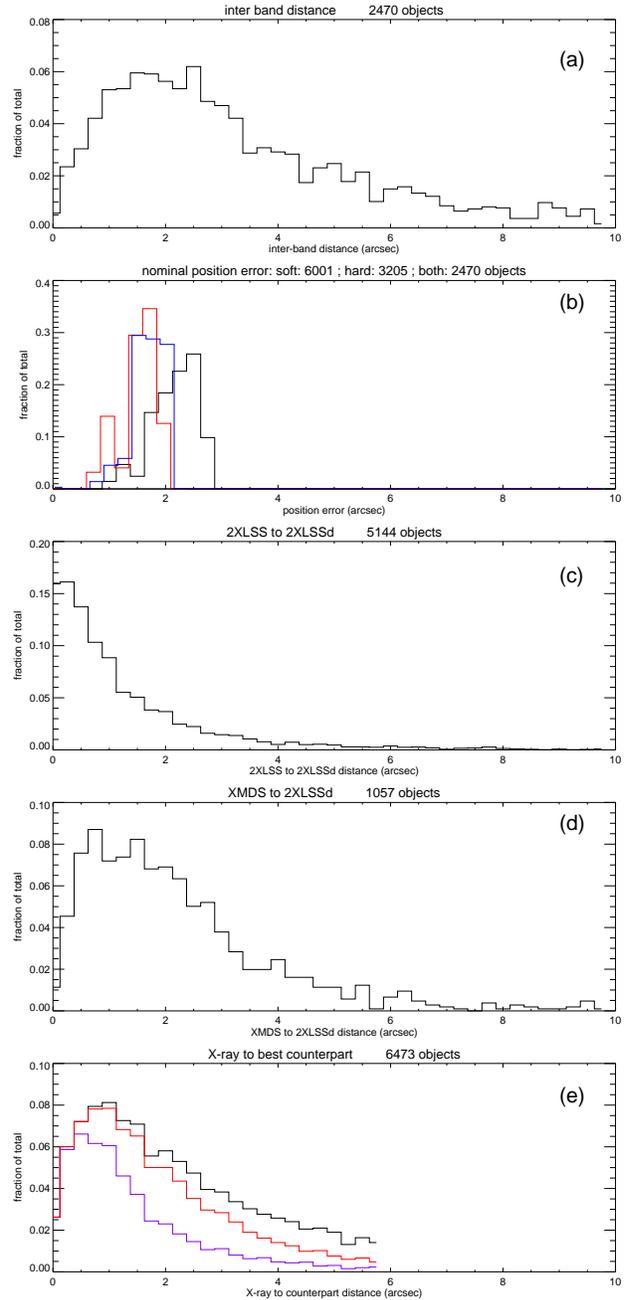} 
   \caption{
   Histograms (as a fraction of the total number of objects indicated in
        each panel title) of distances or positional parameters for the \xlssd~
        catalogue.
        Panel (a) gives the distribution of the inter-band distance {\tt Xmaxdist}
        for objects detected in both energy bands.
        Panel (b) gives the distribution of the position errors for the soft (light gray
        histogram, red in the web version) 
        and hard (light gray histogram, blue in the web version) bands, and of their combined error
        for sources detected in both bands (black histogram).
        Panel (c) gives the distance between the X-ray positions in the two
        catalogues for objects common to \xlssd~ and \xlssII~ (both resulting
        from same \xamin~ pipeline).
        Panel (d) gives the distance between the X-ray positions in the two
        catalogues for objects common to \xlssd~ and XMDS (different event file
        reduction and different pipeline, see Appendix~\ref{SecAPPXMDS}).
        Panel (e) gives the distribution of the distance between the X-ray
        position and the position of the best counterpart in the optical, NIR,
        IR or UV band. The black histogram is for all the objects with a
        counterpart (of any quality) in at least one non-X-ray band. Also shown
        (\textit{as a fraction of the same total}) the distributions for the
        counterparts having either good or fair probability (light gray histogram, red in the web version),
        and for those having good probability (gray histogram, violet in the web version).
        }  
    \label{FigDistHisto}
    \end{figure}

For sources detected in both energy bands, the inter-band distance {\tt Xmaxdist}
is an additional indicator besides the nominal position error described in
Section~\ref{SecAstro}. Its distribution is reported in panel (a) of
Fig.~\ref{FigDistHisto}. If we compute a statistical position error $\sigma$ combining
quadratically the nominal errors in the two bands, we can also see that for \xlssd~
38\% have {\tt Xmaxdist}$\leq \sigma$, 76\% within $2 \sigma$ and 93\% within $3 \sigma$
(for \xlssII~ the percentages are 30\%, 67\% and 88\% respectively).

\subsection{Removal of redundant sources}
\label{SecOverlap}

As in Paper I,  in the case of redundant objects detected in the regions
where the pointings overlap, we keep in the catalogue only the
detection pertaining to the pointing where the source is the
closest to the optical centre (columns ${\tt Boffaxis,~CDoffaxis}$ in the database).
Since overlap removal is the final stage of catalogue building, it is here that
sources  with $LH<15$ are discarded and
only non-spurious sources are brought forward.
However, at variance with Paper I, for analogy with the band merging,
redundant objects are associated within a larger radius of 10\arcsec. Moreover,
the off-axis angle criterion is applied only if the overlapping pointings
are both flagged good or both bad, otherwise the source in the good pointing
prevails unconditionally.

The overlap removal affects 1574 entries in \xlssd~ and 1205 in \xlssII.

Note that the present catalogue also contains a few sources in fields flagged bad.
An extremely conservative usage may exclude all sources detected in bad fields
using condition ${\tt Xbadfield}=0$. A less conservative one should include
the 4 (bad, non-reobserved) fields mentioned in the caption of Table~\ref{TabPointing},

\subsection{Source naming}
\label{SecNaming}

Application of the latest \xamin~ version, of the updated CHFTLS T004 astrometric
corrections,  and of the 10\arcsec~ radius in the
band merging and overlap removal stages, implies that, even in the pointings already
covered by the Version I \xlss~ catalogue, a source may be sometimes superseded by a
different choice, and anyhow may have slightly different coordinates. The same applies
to the two processings (full exposures and 10 ks exposures).
This, combined with the IAU requirement that once a source in a catalogue has been assigned
a name (even if this is a "coordinate name"), the name cannot change even if
the actual coordinates are improved (modified), unless a completely new catalogue is
issued, lead us to define the following naming convention:

\begin{enumerate}
\item the "official" catalogue name ${\tt Xcatname}$ is now generated in the form
      {\tt 2XLSS Jhhmmss.s-ddmmss}, or respectively {\tt 2XLSSd Jhhmmss.s-ddmmss}
      where, as in Paper I, the coordinates used in assigning the name are the ones 
      deduced after the rigid astrometric correction, and chosen as official, i.e.
      those for the {\it best band}  (see Table~\ref{TabMerge}).
\item the single-band catalogue names ${\tt Bcatname}$ and ${\tt CDcatname}$
      use the unofficial
      prefixes {\tt 2XLSSB} or {\tt 2XLSSCD} for both the deep and 10 ks catalogues.
      However, as in Paper I, the coordinates used in the name  correspond to the
      extended (E) or point-like (P) fit in the relevant band (Table~\ref{TabMerge}). 
\item the reference to the \xlss~ source replaced by a \xlssII~ or \xlssd~ source is
      possible using column ${\tt Xlsspointer}$ which contains the value of
      ${\tt Xseq}$ in the table \xlss~ (an explicit look-up in the latter table is
      necessary to find its name or other characteristics). 
\item Similarly, when accessing \xlssII~ it is possible to use column ${\tt Xdeep}$ 
      which points to the value of the ${\tt Xseq}$ closest source in \xlssd.
\end{enumerate}

As described above in Section~\ref{SecMerging}, in a small number of cases, a  source
in a band happens to be associated  with two different objects in the other
band. These couples of catalogue entries are flagged by  a non-zero value in column 
${\tt Xlink}$,
Consistently with the convention defined in Paper I,
the ambiguity in the name is resolved (when necessary, i.e. in 1 case for \xlssd~
and 3 cases in \xlssII)
by the addition of a suffix:
e.g. the two members of a couple will appear as {\tt 2XLSS
JHHMMSS.S-DDMMSSa} and {\tt 2XLSS JHHMMSS.S-DDMMSSb}.

\subsection{Extended source classification}
\label{SecC1C2}

The extended source classification is the same as described in Paper I.
Extended sources are selected from the \xamin~ parameter space
as detections with {\tt extent}~$ > 5\arcsec$,  {\tt likelihood of extent}~$>15$,
and further divided into two classes: C1 with {\tt likelihood of extent}~$>33$ 
and {\tt likelihood of detection}~$>32$, which is almost uncontaminated by misclassified
point sources, and C2 (the rest), allowing for $\approx 50\%$ contamination.
This classification is rather stable even in case of changes in the exposure time or background,
as shown in Fig.~9 of \cite{2012MNRAS.423.3561C}

The catalogues report only the flagging as extended source in the soft band
(column ${\tt Bc1c2}$). However, for the unique purpose of band merging,
the same classification has nominally been applied also to the hard band.
A short statistics is reported in Table~\ref{TabWasTabularExt}, while
the compatibility in the two catalogues (\textit{compatible} means extended in
both catalogues in the (prevailing) band where it is detected, and undetected (or
point-like) in the other band) is shown by this breakdown:
\begin{description}
\item 95 sources with same extended classification
\item  8 sources with compatible classification
\item  1 soft extended in \xlssII, hard extended in \xlssd
\item 23 point-like in \xlssII, extended in \xlssd
\item 20 extended in \xlssII, point-like in \xlssd
\end{description}

Studies of galaxy clusters in the \xmm-LSS are presented in
\cite{2011A&A...526A..18A}, 
 \cite{2012Willisinprep} and
\cite{2012Clercinprep}.
More information on confirmed clusters is also available in the \xmm-LSS survey
cluster database\footnote{\url{http://xmm-lss.in2p3.fr:8080/l3sdb/}}.

\begin{table}
\caption{Statistics of extended sources}
\label{TabWasTabularExt}
\begin{center}
\begin{tabular}{lrr}
\hline
 Classification and catalogue & \xlssII & \xlssd \\
\hline\hline
sources with ${\tt Bc1c2} \ne 0$             &  147 &  160 \\
~~of which C1                                &   49 &   53 \\
~~of which C2                                &   98 &  107 \\
total no. of extended sources                &  187 &  210 \\
~~of which C1                                &   54 &   57 \\
~~of which C2                                &  133 &  153 \\
~~extended in both bands                     &    4 &    4 \\
~~soft C1 detected in both bands             &    9 &   12 \\
~~soft C2 detected in both bands             &   12 &   16 \\
~~soft only C1                               &   36 &   37 \\
~~soft only C2                               &   86 &   91 \\
~~\textit{hard only C1}                      &    5 &    4 \\
~~\textit{hard only C2}                      &   35 &   46 \\
\hline
\end{tabular}
\end{center}
\end{table}

\section{Generation of the multiwavelength catalogue}
\label{SecOpt}

\subsection{The input catalogues}
\label{SecOptInput}

Most of the \xmm-LSS area was covered in the $u^{*},g',r',i',z'$ bands
by the W1 Wide Synoptic field of the CFHTLS, and the core area
(our G pointings in Table~\ref{TabPointing}, corresponding to the XMDS) also by the 
1 \dd~ Deep field D1. The northernmost strip $\delta \ga -3.7\degr$ was
not part of CFHTLS and was observed under a Guest Observer program
in a similar configuration at the CFHT (but only in the $g',r',z'$ bands)
with three pointings (so called ABC fields), leaving a gap only in correspondence
of the bright star Mira Ceti.

We used a compilation of the Terapix\footnote{\url{http://terapix.iap.fr/}}
panchromatic catalogues for W1 (release T004) and the
ABC fields, edited to get rid of duplicates in overlapping pointings, 
and replacing undefined magnitudes due to non-detection in one band with
the limiting magnitude of the pointing.
Separately we also used the panchromatic catalogue of the D1 field.

Most of the \xmm-LSS area was also observed by the \textit{Spitzer Space Telescope},
as part of the SWIRE survey \citep{2003PASP..115..897L}. We obtained
from IPAC a compilation of an unpublished release 
catalogue in the 4 IRAC (3.6, 4.5, 5.8 and 8.0 $\mu$m)
and 3 MIPS (24, 70 and 160 $\mu$m) bands, which was pre-processed for 
classification of extended objects; in  particular IRAC fluxes are Kron fluxes for
extended objects and so-called aperture 2 (1.9\arcsec) otherwise, while they are
APEX (PRF) fluxes for MIPS.

For the UKIDSS NIR survey we retrieved from the WSA public archive\footnote{\url{http://surveys.roe.ac.uk/wsa/}}
data (within 10\arcsec~ from our X-ray source positions)
from the release DR5plus, which at the time provided partial coverage of some areas
of the \xmm-LSS via the DXS and UDS surveys
(in particular the latter covers the SXDS area).

For the UV band we retrieved from the NASA MAST public \textit{GALEX} archive\footnote{\url{http://galex.stsci.edu/GR4/}}
(using the {\sc CasJobs} tool) data from the GR4/GR5 release
(within 10\arcsec~ from our X-ray source positions).
Since it is well known that the MAST \textit{GALEX} catalogue contains redundant sources where
\textit{GALEX} pointings overlap (so called \textit{tiling artifacts}), we have run a procedure
to flag \textit{GALEX} objects within 1.5\arcsec~ from any other one observed in a different tile,
and to prefer in each set the one observed in two bands, or with smallest inter-band separation,
or with smallest off-axis angle.

The coverage of the \xmm-LSS area by the various catalogues is shown in
Fig.~\ref{FigPointing}.

All appropriate data were ingested in tables within our database and elaborated
therein with the procedure described below (the optical data were also used separately
for the \textit{preventive}
astrometric correction described in Section~\ref{SecAstro}).

\subsection{The candidate definition procedure}
\label{SecOptIdent}

As a preliminary step, we construct within our database correlation tables between
the X-ray sources in either \xlssII~ or \xlssd~ and \textit{each one independently}
of the CFHTLS D1, W1, SWIRE, UKIDSS and \textit{GALEX} tables, using a radius of
6\arcsec.

The next step is an \textit{incremental addition} procedure through the above tables
in the quoted order:

\begin{enumerate}
\item
We create a generalized correlation table, with columns designated to hold
 pointers to the various catalogues, initialized with as many records as X-ray sources.
 Each record is an n-uple $(X,D1,W1,SWIRE,UKIDSS,GALEX)$ initialized as
 $(X,0,0,0,0,0)$.
 Records are termed \textit{counterpart sets}.
\item
 We start with the D1 table and for each X-ray source we
 {\it insert a pointer} in the relevant record
 if there is at least one D1 object
 within 6\arcsec. If the X-ray source has one optical counterpart only, the D1 pointer
 is {\it inserted} in the existing primary record,
 so the n-uple is filled as $(x_1,d_a,0,0,0,0)$.

 If it has more, 
 the pointer of the closest candidate is inserted as above, while
 {\it additional records are added} copying from the primary one and
 replacing the pointer with one associated with the other D1 object,
 e.g. an additional n-uple $(x_1,d_b,0,0,0,0)$.
\item
 Then one proceeds in turn to the next table \textit{inserting}
 an object from such a table when it is closer to one of the existing counterparts
 in other non-X-ray tables within a predefined radius.
      Only objects within 6\arcsec~ from the relevant X-ray source are considered, while a
      correlation radius of 0.5\arcsec~ is used when comparing positions of
      the same origin (i.e. D1 and W1), of
      1\arcsec~ when comparing to SWIRE or UKIDSS catalogues,
      and of 1.5\arcsec~ when comparing to GALEX.
\item 
 A pointer is {\it inserted} in an existing record when there
 is a single match with the X-ray position and all the positions in the
 previously processed catalogues.
 E.g. an n-uple is updated as $(x_1,d_a,w_a,0,0,0)$
 {\it Additional counterpart sets are generated} in all other cases
 (typically an independent counterpart of the X-ray source with no
  counterpart in previous catalogues, but could also be an ambiguous association
  of more sources in the current catalogue with a previously defined counterpart
  set).
  E.g. in the case of W1 objects they are compared with D1, while
  SWIRE objects are compared first with W1, then D1;
  UKIDSS objects are compared with preceding tables (in order W1, D1, SWIRE);
  and GALEX objects are compared with all other tables (in order W1, D1, SWIRE,
  UKIDSS). One may therefore end up with completely or partially filled pre-existing
  n-uples like   $(x_1,d_a,w_a,s_a,0,g_a)$, or with new n-uples like
  $(x_1,0,w_n,s_n,u_n,0)$ or $(x_1,0,0,s_p,0,0)$, or (seldom) with ambiguous cases like
  $(x_1,d_b,w_b,0,0,g_q)$ and $(x_1,d_b,w_b,0,0,g_r)$.
\item Finally the chance probabilities for random association of a counterpart with
      the X-ray source are computed as described immediately hereafter.

\end{enumerate}

\subsection{Computing probabilities and counterpart ranking}
\label{SecOptProb}

We compute
{\it four} probabilities: ${\tt probXO}$, ${\tt probXS}$,  ${\tt probXU}$
and ${\tt probXG}$; each is
the probability of chance coincidence between the X-ray source and its
counterpart in a given catalogue, based on the X-ray to optical (or SWIRE, UKIDSS or \textit{GALEX}) distance,
the optical, IR or UV intensity (magnitude or flux) $m$, and the density of sources 
brigther than such an intensity.
They are based on a formula 
\citep{1986MNRAS.218...31D}
like

\vskip 5pt
\begin{equation}\label{Eq1}
  prob = 1 - exp(-\pi~ n(brighter~than~m)~  r^2 )
\end{equation}
\vskip 5pt
~
where $r$ is the X-ray to counterpart distance, while
a rough estimate of
the density $n(brighter~than~m)$ is computed as described in detail
in Appendix \ref{SecAPPOptProb}.

At this stage each X-ray source can have more than one potential counterpart (or
better, counterpart sets, where each set may include associated counterparts in D1,
W1, SWIRE, UKIDSS and \textit{GALEX}).
A preliminary ranking can be assigned on coarse probability ranges:
\begin{description}
\item good if $prob < 0.01$
\item fair if $0.01 < prob < 0.03$
\item bad if $prob > 0.03$ 
\end{description}

The acceptable tuning with the data of such coarse classification is 
demonstrated by Fig.~\ref{FigHisto}.
Such a pre-ranking is refined by a multi-step heuristic procedure, which assigns
a score based on several criteria (for instance weighing more a good or
fair probability in the optical or SWIRE bands, or the fact that the best
probability of a counterpart set is at least 10 times better (smaller)
than those of any other counterpart set for the same X-ray source, or
whether the counterpart set is unique, or brightest and closest).
In some cases a visual inspection of the optical thumbnail (see Section~\ref{SecODP})
with the overlay of all counterpart set elements has been necessary. 
In exceptional cases this resulted in a manual editing (usually deletion of
counterpart sets due to artifacts, like unresolved tiling effects in one of
the catalogues, or problems near very bright or saturated sources).

The result of the ranking is the assignment of the value of column ${\tt Xrank}$
(see Table~\ref{TabMulti})
in the multiwavelength catalogues {\tt 2XLSSOPT} or {\tt 2XLSSOPTd} (derived respectively from
the 10 ks and deep X-ray catalogues).

The number of \textit{potential} counterpart sets can be rather high (16813
for \texttt{2XLSSOPT} and 20837 for \texttt{2XLSSOPTd}). However a large
number of them (9093 and 11500), based
on the above ranking procedure, obtain a rank
${\tt Xrank=-1}$, which means they have to be rejected.
This leaves 7720 or resp. 9337 potential counterpart sets in the publicly released
\texttt{2XLSSOPT} or \texttt{2XLSSOPTd}. 
Such non-rejected counterpart sets, as the result of an ambiguity analysis, have
${\tt Xrank}$ between 0 and 2. For X-ray sources with a single counterpart set
(either physically unique or just one non-rejected) ${\tt Xrank}$ is either 0 or 1.
When instead an X-ray source has more possible counterpart
sets, there is a single one which has ${\tt Xrank=0}$ or ${\tt Xrank=1}$, i.e. the
preferred, while all the secondaries have ${\tt Xrank=2}$. 
More details on ranks, together with a
detailed statistics, are presented in Section~\ref{SecOStat1}.
The rank, and the potential counterpart list, are
provided as a convenience for database users, but are not at all
intended as prescriptive.
Additional information about visual, spectroscopic and SED classification of X-ray
sources with respect to the optical counterparts may be found in \cite{2012Melnykinprep}.

   \begin{figure*}
   \centering
   \includegraphics[width=8.5cm]{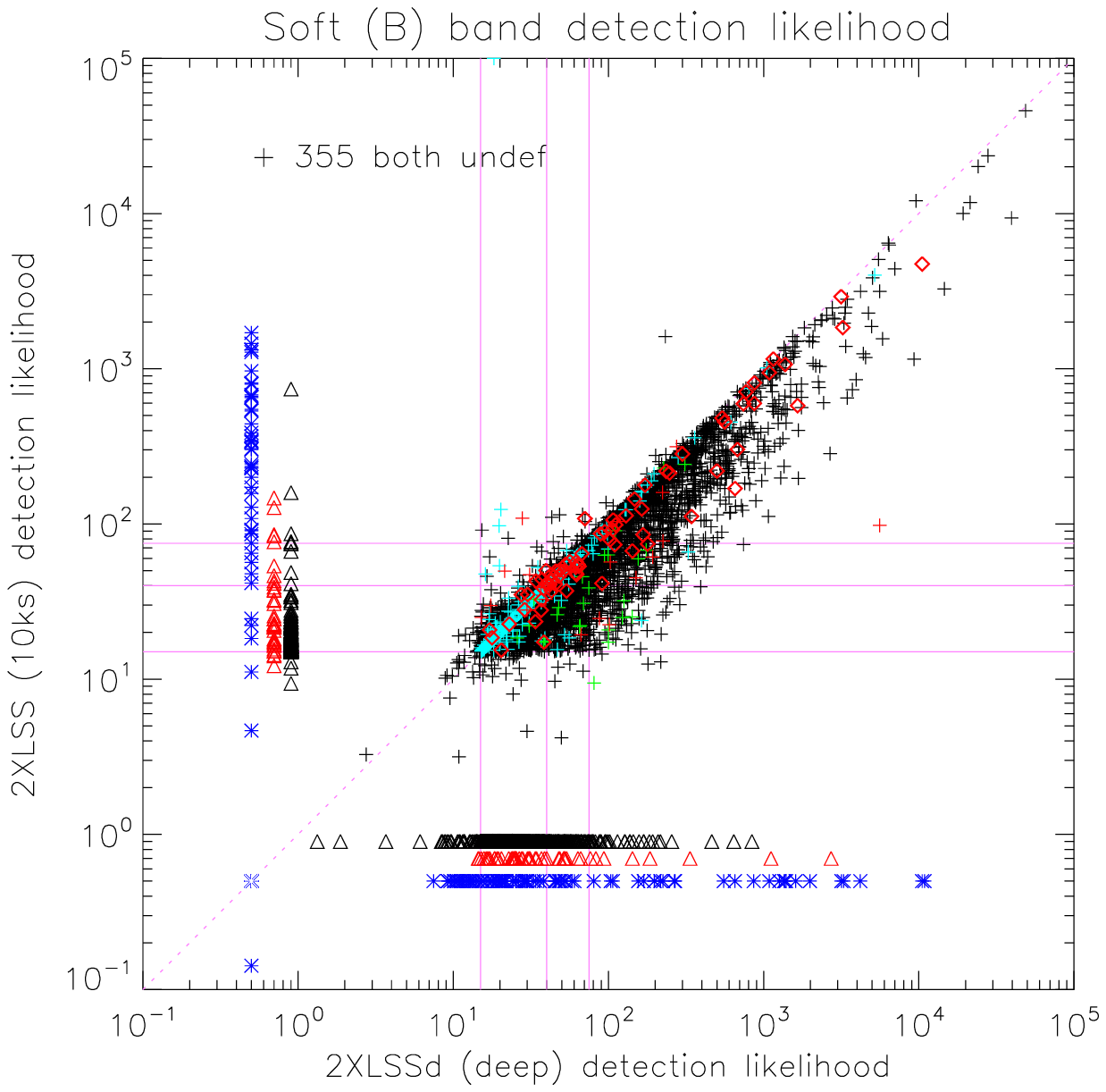} 
   \includegraphics[width=8.5cm]{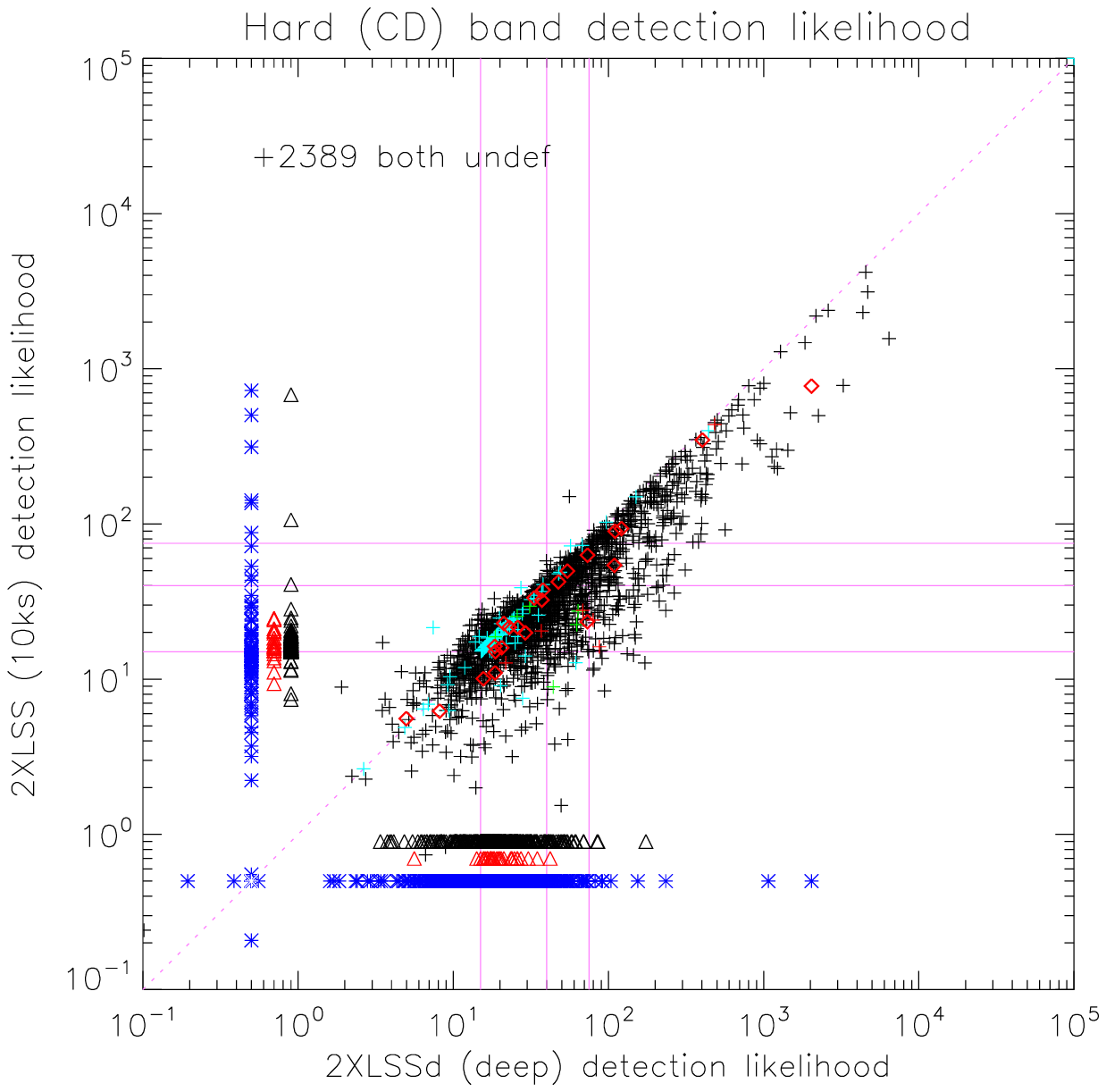} 
   \includegraphics[width=8.5cm]{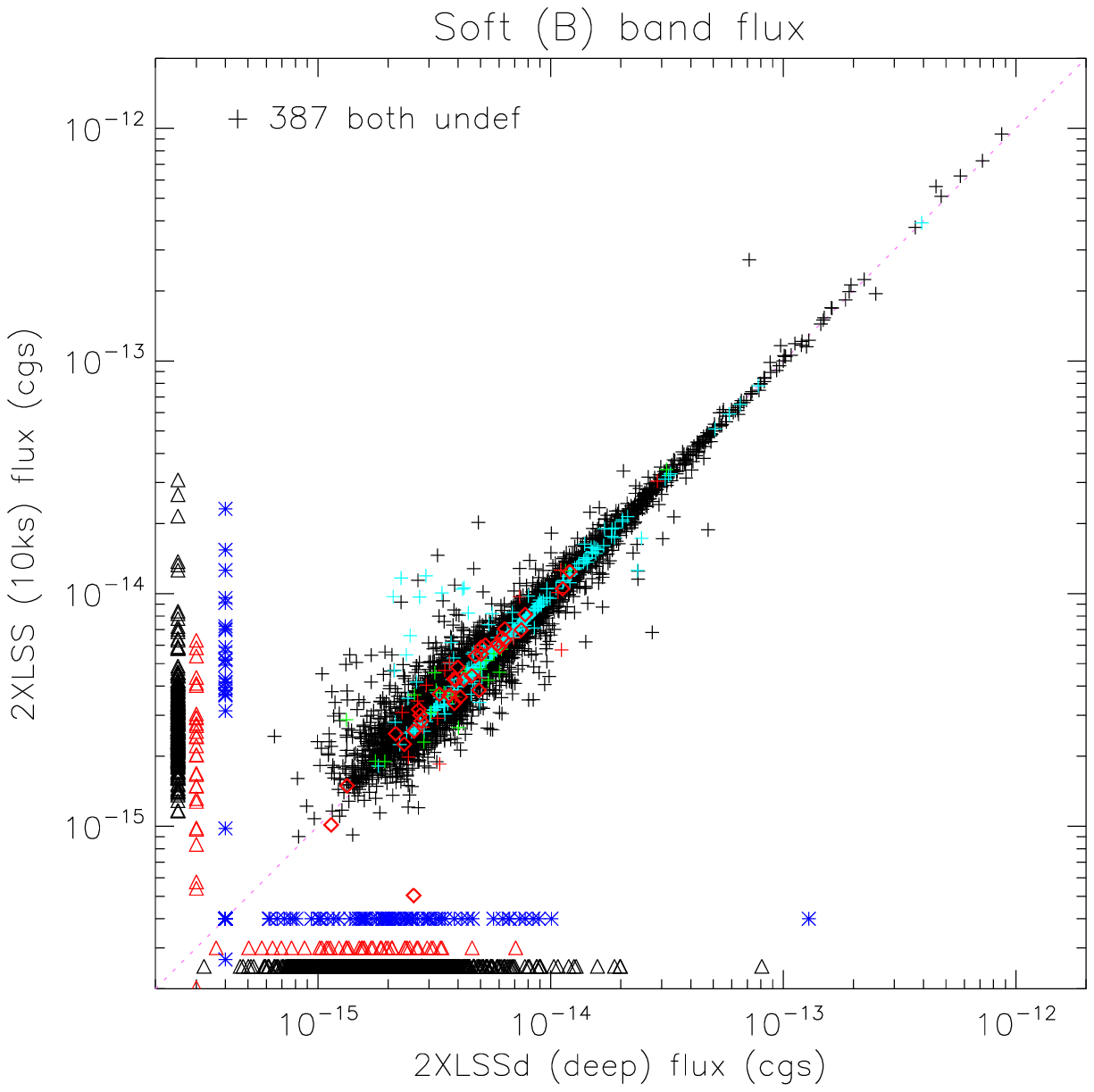} 
   \includegraphics[width=8.5cm]{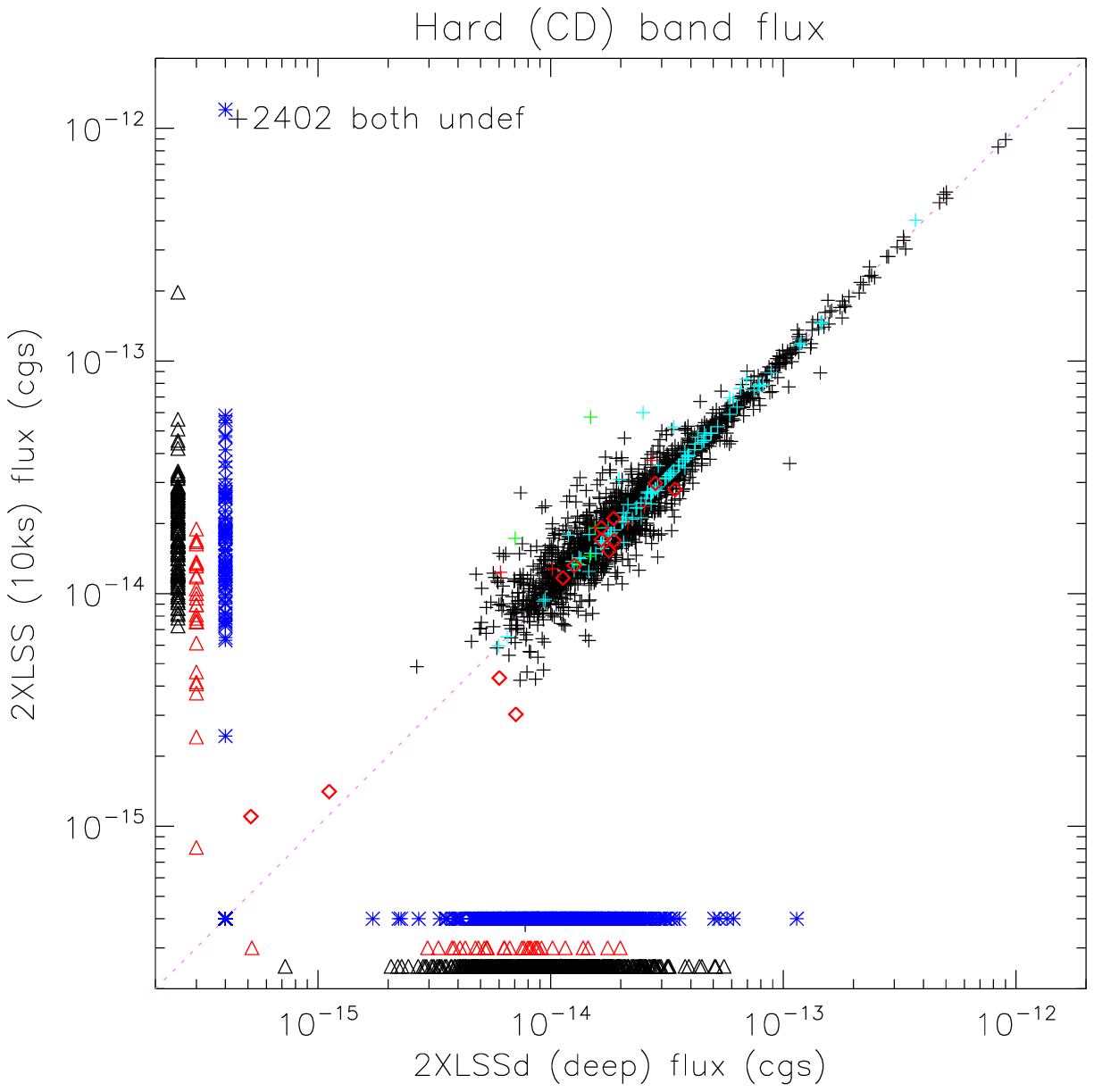} 
   \caption{Comparison of the detection likelihood (top row)
            and of the flux (bottom row)
            in the soft (left column) and hard (right column) energy bands
            between \xlssII~ and \xlssd.
            Crosses and diamonds indicate point-like or extended objects
            associated in the two catalogues (see text).
            Blue asterisks indicate likelihood or flux are present but
            \textit{undefined} in one catalogue,
            while triangles indicate sources present only in one catalogue
            (both are placed at a conventional out-of-range X or Y position).
            The number of objects with undefined values in \textit{both}
            catalogues in a given band, but nevertheless associated, is
            indicated near the top left corner of each panel.
            Colour coding (only in the web version) is as follows:
            black cross for point-like common sources in \xlssII~ good fields;
            cyan cross idem for bad fields; green cross for \xlssd~ extended
            object point-like in \xlssII; vice versa for red cross; red diamond
            for extended sources in both \xlssII~ and \xlssd.
            Triangles are black or red for point-like or extended sources which
            are either new in \xlssII~ or present in \xlssd~ but
            lost in the shallower catalogue.
            In the likelihood plots, the thin pink lines are fiducial marks
            corresponding to the spurious/non-spurious threshold (15) and to
            the conventional $3 \sigma$ (40) and $4 \sigma$ (75) levels.
             }
   \label{FigCompa2}
    \end{figure*}

\section{A few statistics}            
\label{SecStat}

\subsection{The X-ray catalogues}     
\label{SecXstat}

The number of sources in the merged 10ks catalogue is
5572 for \xlssII~ (4932 in {\tt 2XLSSB} and 1923 in {\tt 2XLSSCD}).
The number of sources in the deep catalogue is
6721 for \xlssd~ (5881 in {\tt 2XLSSBd} and 2645 in {\tt 2XLSSCDd}).

A majority of objects detected in the full exposures are
confirmed in the 10 ks exposures, 
usually with the same classification and within a distance of 6\arcsec;
the differences are concentrated within the objects with poorer likelihood.
However there is a significant number of detections, not necessarily spurious,
which are either present only in the full exposures (not surprising) or
even only in the 10ks exposures.
If one considers data \textit{before} band merging \textit{and} spurious source filtering,
24\% of the soft detections and 40\% of the hard detections in full exposures
are not confirmed in 10 ks ones, while 8\% (soft) and 19\% (hard) 10 ks detections 
are new.

\begin{table}
\caption{Basic statistics for the \xlssII~ and \xlssd~ catalogues}
\label{TabWasTabular}
\begin{tabular} {lrr}
Case                               & \xlssd & \xlssII \\
\hline\hline
Total number of sources            & 6721 & 5572 \\
\hline
Detected in both bands                           \\
~~non-spurious in both bands       & 27\% & 23\% \\
~~non-spurious in soft band only   &  8\% &  8\% \\
~~non-spurious in hard band only   &  2\% &  1\% \\
Detected only in soft band         & 52\% & 57\% \\
Detected only in hard band         & 11\% & 10\% \\
\hline
Not detected in other catalogue & 24\% & 9\% \\
\hline
\end{tabular}
\end{table}

Considering band merged data \textit{before} overlap removal (and spurious source 
filtering), 29\% of the merged detections in full exposures are not confirmed in 10 ks 
ones (mainly single soft non-spurious, or in similar proportion between hard and soft when spurious),
while 14\% of the 10 ks detections are new (the majority are spurious, but mainly single
detections in the soft band prevail when non-spurious).

We proceed below to some further comparison between the deep and 10 ks catalogues,
which allows us to assess a trade-off between deeper but disuniform exposures and
shallower uniform exposures.
A comparison with the Version I (\xlss) release is reported in
Appendix \ref{SecXstatNewOld}.

While sources in a \textit{catalogue} are by construction non-spurious (i.e.
with $LH>15$ \textit{in at least one band}), they can
be detected as such in both bands, detected as non-spurious in one band and spurious
in the other, or detected in a single band. 
The breakdown in percentage is reported in Table~\ref{TabWasTabular}.
The deep catalogue is marginally better for what concerns full-fledged
both band detections.

A breakdown considering also the classification for the 5117 common objects
can be summarized as follows: 85\% are classified point-like identically in both
catalogues (which means either point-like or undetected in each band), which goes up to 
97\% classified point-like and compatible (i.e. detected in both bands in one catalogue
and in a single one in the other); 2\% are classified as extended (usually identically,
only compatible in 8 cases, while in 1 single case the source is detected as extended
once in the soft and once in the hard band); the few remaining cases are 23 \xlssd~
and 20 \xlssII~ extended sources which are point-like in the other catalogue.

The distances between common objects are in very good agreement:
90\% within 2\arcsec, 97\% within 4\arcsec, and 99\% within 6\arcsec
(see also panel (c) of Fig.~\ref{FigDistHisto}).

A comparison of likelihoods and fluxes for sources associated
is reported in Fig.~\ref{FigCompa2}. As expected the likelihood in the shorter
exposure \xlssII~ catalogue is compatible but lower than the one in \xlssd~ (the
points lay below the diagonal fiducial line of equal values).

For fluxes they are generally rather well consistent (with exceptions for a few
extended sources), with only a moderate scatter for fainter objects.
In lack of error bars, one can compare the compatibility of fluxes for the 5144
sources associated between \xlssd~ and \xlssII. For the 4656 with a soft-band
detection in both catalogues, 60\%, 81\% and 96\% of the sources have fluxes
within 10\%, 20\% or 50\%. The equivalent percentages for the 2146 with a hard-band
detection are 53\%, 77\% and 95\%.

\begin{figure}
   \centering
   \includegraphics[width=8.4cm]{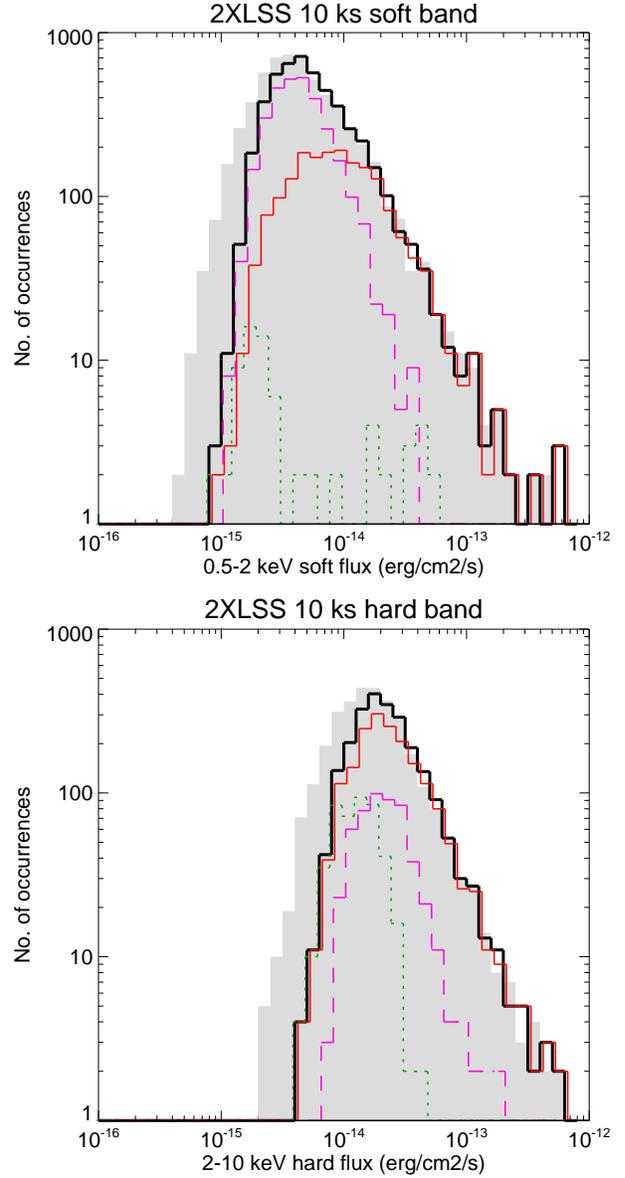} 
   \caption{Histogram of the soft (top panel) and hard (bottom panel) fluxes.
            The gray shaded histogram in the background refers to the \xlssd~ deep
            catalogue. All other histograms refer to the \xlssII~ 10 ks catalogue,
            and namely to: all sources in the band (thick black); sources
            detected in both bands (thin solid red);  sources detected only in
            the band (hence by definition non-spurious; thin dashed magenta);
            fluxes with a spurious likelihood in the band, associated with a
            non-spurious source in the \textit{other} band (thin dotted green).
            Colours are shown only in the web version.
            The various histograms are slightly offset for clarity.
            }
    \label{FigFluxHisto}
    \end{figure}

In Fig.~\ref{FigFluxHisto} we provide also an histogram of the fluxes, mainly for 
the \xlssII~ catalogue (but the shaded area indicates what we "gain" at low fluxes passing
from 10 ks to full exposures).

\subsection{The multiwavelength catalogues} 
\label{SecOstat}

\subsubsection{Statistics on each catalogue \label{SecOStat1}} 
~\newline

We first present some general statistics on both catalogues 
in parallel, quoting values for {\tt 2XLSSOPT}, followed by
those for {\tt 2XLSSOPTd} in parentheses.

   \begin{figure}
   \centering
   \includegraphics[width=8.4cm]{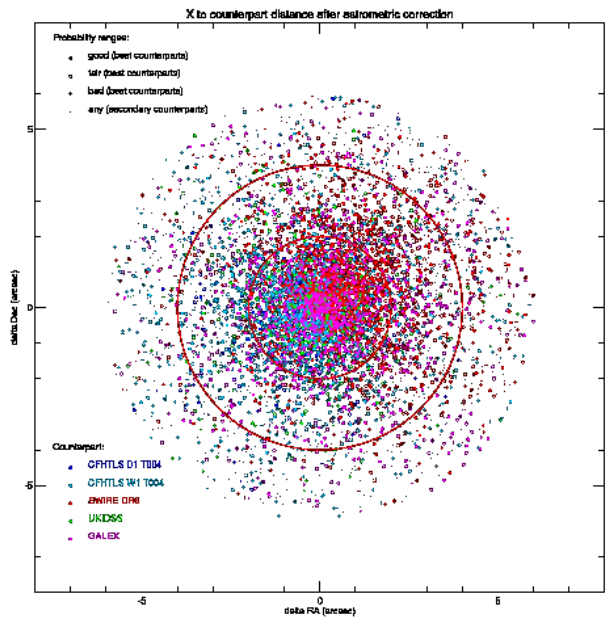} 
   \caption{Distances in RA and Dec between the X-ray corrected position and
            the counterpart position. Different symbols indicate the
            identification quality. 
            A circle is plotted when the counterpart is the best one, and
            the chance probability is good or fair (filled in case of good probability).
            A cross is plotted for the best counterpart when the probability is bad.
            A dot is plotted for secondary (ambiguous) counterparts, irrespective of
            probability, but only if it is good or fair.
            Different colours (web version only) or
            shades (as shown on figure) indicate the origin
            of the counterpart position for the distance calculation.
            Two fiducial radii of 2 and 4\arcsec\  are also shown.
            This figure refers to \xlssII; the equivalent figure for \xlssd~
            is extremely similar.
            }
    \label{FigAstroDist}
    \end{figure}

To evaluate whether in a given region we do not find counterparts
in a given wavelength table because they do not exist or because the region has not been 
observed at all, one should refer to Fig.~\ref{FigPointing}.

{\tt 2XLSSOPT} ({\tt 2XLSSOPTd}) starts from 16813 (20837) nominal counterpart sets,
from which we removed the 9093 (11500) rejected (${\tt Xrank=-1}$ as explained
in Section~\ref{SecOptProb}).

X-ray sources nominally flagged as {\it blank fields} (i.e. having a single 
\textit{null counterpart set}, i.e. no catalogued CFHTLS,
SWIRE, UKIDSS or GALEX counterpart within 6\arcsec) are 221 (248).
Note that the absence of catalogued sources does not mean they are necessarily 
real blank fields. Often bright sources are omitted by the catalogues, but are visible
if one inspects the thumbnail image. Compare for instance the cases of sources
${\tt 2XLSSOPT.Xseq=43302}$, which is very close to a R=15.6 galaxy shown in SIMBAD, or
${\tt 2XLSSOPT.Xseq=38678}$ whose field is spoiled by the nearby bright star BD-05 427.
So some of the cases flagged as blank field can instead have a bright counterpart.

Concerning tentative identifications of 5572 (6721) X-ray sources:
\begin{description}
\item
18\% (17\%) have a \textit{physically single counterpart} (set)
\item 
40\% (39\%)
have a \textit{single very reliable counterpart}, i.e.
      ${\tt Xrank=0}$ plus eventual \textit{rejected} counterpart sets
\item 
21\% (28\%)
have a \textit{single , but not so reliable, counterpart}, i.e.
      ${\tt Xrank=1}$, also plus eventual \textit{rejected} counterpart sets
\item  
16\% (16\%)
are \textit{pseudo-ambiguous}, with one
      \textit{definitely} preferred counterpart (${\tt Xrank=0}$), plus one or more
      nominal secondary counterparts with rank 2.
\item  
13\% (14\%)
are \textit{definitely ambiguous}, with one
      \textit{nominally} preferred counterpart (${\tt Xrank=1}$), plus one or more
      secondary counterparts with rank 2, at least one of which is not terribly
      worse than the nominally preferred one.
\end{description}

With reference to the criteria defined in Section \ref{SecOptProb},
48.2\% (48.6\%) of the sources have a best counterpart with a good probability,
29.5\% (30.1\%) with a fair one, and 4.0\% (3.7\%) are nominal blank fields.

One might also relate the counterpart association with the X-ray detection
significance (using the
cross calibration between likelihood and number of $\sigma$ presented in
Appendix~\ref{SecAPPXMDSX}): for instance of
1412 (1888) X-ray sources detected above $4\sigma$, 
77\% (76\%) have a good counterpart, 
18\% (20\%) a fair one, and only
2\% (1\%) are unidentified;
of 2436 (3169) X-ray sources  above $3\sigma$, 
68\% (65\%) have a good counterpart,
25\% (27\%) a fair one,
and 2\% (2\%) are unidentified.

One shall also note that the \textit{ranking depends} on the probabilities, and these 
depend on the distance (Section~\ref{SecOptProb}) and therefore \textit{ultimately on 
the X-ray position}. If the latter changes, the rank choice will change. 
The differences between the two catalogue variants are discussed in the next section.

Finally, the quality of the tentative identifications can be assessed from the
offset (distance) between the X-ray source position and the position of the best counterpart
in the best counterpart set. This is shown in Fig.~\ref{FigAstroDist}
and in panel (e) of Fig.~\ref{FigDistHisto}.
  83\% (84\%) of all counterparts have a distance within 4\arcsec, which 
  occurs for 90\% (91\%) of the best counterparts with fair or good
  probabilities (the circles in Fig.~\ref{FigAstroDist}) and for
  95\% (95\%) of those with good probabilities 
  (the filled circles in Fig.~\ref{FigAstroDist}).

There is some evidence from Fig.~\ref{FigAstroDist} of a systematics in the
deviations between X-ray positions and positions in the various catalogues.
The average deviation for the optical and UKIDSS catalogues clusters around a
point in the third quadrant (e.g.$ -0.39\arcsec,-0.07\arcsec$ for W1),
while the one for SWIRE
clusters around a point in the first quadrant ($0.82\arcsec,0.57\arcsec$).
For \xlssd~ ($-0.40\arcsec,-0.07\arcsec$) and ($0.79\arcsec,0.52\arcsec$) respectively.

Concerning panel (e) of Fig.~\ref{FigDistHisto}, it reports the distribution
of the X-ray to counterpart distance, using as counterpart position the one
with the smallest distance
(in 37\% this is an optical source, in 34\% a SWIRE one, in 9\% a UKIDSS one and
in 20\% a \textit{GALEX} one).

\subsubsection{Differences between {\tt 2XLSSOPT} and {\tt 2XLSSOPTd} \label{SecCompOct}} 
~\newline

The differences between the two catalogues with optical identifications derive from
three main reasons, the former two physiological, due to the different exposures:

\begin{enumerate}
\item the X-ray source may be detected in one of the input catalogues and not
      in the other 
\item the X-ray source may be detected or classified differently (spurious or
      non-spurious, in one or two bands, point-like or extended)
\item the X-ray source can be detected at a displaced position
\end{enumerate}

The latter displacement may result in some of the possible counterparts be outside
the 6\arcsec~ correlation radius, and therefore in the list of counterpart sets being
partially or totally different, and with different ranks.

Considering all potential counterpart sets (including negative rank \textit{rejected}
counterparts since rejection may act differently because of the displacement), about
86\% of the common ones are identical (i.e. have the same counterparts in all non-X-ray catalogues),
of which 139 are confirmed "blank fields" (no \textit{catalogued} counterpart in any waveband).
The remaining cases may be altogether different counterpart sets, or partially match
(in some of the non-X-ray catalogues).
More details are provided in Appendix~\ref{SecAPPCompOct}

Concerning the tentative \textit{blank fields} one has to note that,
besides the 139 common ones, there are 109 {\tt 2XLSSOPTd} blank fields not
present in the 10ks catalogue and  82 {\tt 2XLSSOPT} blank fields which are new in
the latter catalogue (42 are new X-ray sources with no deep counterpart, the other
40 are no longer blank fields).

Coming now to \textit{non} blank fields,
remember that the identification procedure is incremental. So it starts (in absence of
a D1 counterpart) associating a W1 object with the X-ray source. Then it may append one SWIRE
object (associated with the X-ray source and within 1\arcsec from W1) to the counterpart set,
and create a new counterpart set for another SWIRE object. 
And so on and so forth for the other wavebands.
Each association may be different as a result of a small displacement in the X-ray position.
In the most favourable case this may just prefer a particular counterpart in a
counterpart set otherwise identical and identically ranked. 
In other cases counterpart sets similar but differing in one waveband may be
ranked differently (primary vs secondary or even rejected).

\subsection{Comparison with XMDS} 
\label{SecXMDS}

This section provides a sketchy comparison between the catalogues presented in this paper
and the XMDS (\xmm~Medium Deep Survey) one, a subset of which was published as
the XMDS/VVDS $4\sigma$ catalogue (\citealp{2005A&A...439..413C}).
Since the entire XMDS catalogue is unpublished, we will release it through our
database contextually with Version 2 \xmm-LSS ones (see Table~\ref{TabTables}).
~
A comparison provides an opportunity to validate and cross-calibrate two
different pipelines (a traditional and an innovative one) on the same input data.
The main differences between the two pipelines, and further details of the
comparison are reported in Appendix~\ref{SecAPPXMDS}.

The XMDS catalogue includes 1168 sources (by definition all in the G-labelled fields)
of which  1057 are catalogued in \xlssd.
Appendix~\ref{SecAPPXMDSX} provides further details, in particular:
\begin{enumerate}
\item
A cross calibration of the detection likelihood of
\xamin~ with the significance in terms of the number of $\sigma$ of the XMDS shows
that a likelihood of 75 corresponds more or less to the $4\sigma$ level,
and one of 40 to the $3\sigma$ level.
\item
Fluxes match rather well, although with a systematic difference
(which, considered the different procedures is fully acceptable), namely
\xlssd~ fluxes are 
0.895 lower than the XMDS ones in the B band, while they are only
1.040 higher in the CD band.
\item Also XMDS fluxes measured \textit{simultaneously} in all bands match
      well \xlssd~ fluxes measured \textit{separately} for sources detected in
      both bands, which reinforces trust in the band merging procedure described
      in Section \ref{SecMerging}.
\end{enumerate}

It is also possible to compare the
counterparts in optical (and other) bands between the XMDS and the
{\tt 2XLSSOPTd} catalogue 
and, as shown in Appendix~\ref{SecAPPXMDSOpt},
the compatibility between the counterparts is also satisfactory.

\subsection{Comparison with 2XMM} 
\label{SecWatson}

We did a quick comparison with the 2XMM (second \xmm-Newton serendipitous source)
catalogue \citep{2009A&A...493..339W}.
Details are reported in Appendix~\ref{SecAPPWatson}.
We find, despite the differences in the data processing and in the
definition of the energy band, an acceptable match in terms of number of sources,
respective distance, and fluxes.

\section{Concluding remarks}
\label{SecConclusion}

We have presented in this paper X-ray full-exposure (\xlssd) and 10
ks-limited exposure (\xlssII) catalogues for the 11.1 \dd~ \xmm-LSS field.
The total number of X-ray sources reported in these two catalogues are
6721 and 5572, respectively.
The sources were detected in the 0.5-2 keV
and/or 2-10 keV energy bands with a new version (3.2) of the \xamin~ pipeline.
We have also provided two multi-wavelength catalogues (cross-correlating
out the X-ray sources with IR/SWIRE, NIR/UKIDSS, optical/CFHTLS and UV/\textit{GALEX}
sources),
\texttt{2XLSSOPTd} and \texttt{2XLSSOPT},
corresponding to the full and 10 ks-limited exposure catalogues respectively. 
We have also described in detail the
X-ray band merging, the X-ray point-like and extended source classification, the
matching procedure of counterparts from multiwavelength surveys as well as
extensive statistics to compare the two presented catalogues between them and with
previous studies.
 
Catalogues and associated data products are available through the
Milan data base 
(\url{http://cosmosdb.iasf-milano.inaf.it/XMM-LSS/}),
with a reduced summary stored at CDS.

\section*{Acknowledgments}

The results presented here are based on observations obtained with
\xmmn, an ESA science mission with instruments and
contributions directly funded by ESA Member States and NASA.
\\
We acknowledge the work done by Krys Libbrecht on the \xamin~ pipeline
translation.
\\
Optical photometry data were obtained with MegaPrime/MegaCam, a
joint project of CFHT and CEA/DAPNIA, at the Canada-France-Hawaii
Telescope (CFHT) which is operated by the National Research
Council (NRC) of Canada, the Institut National des Sciences de
l'Univers of the Centre National de la Recherche Scientifique
(CNRS) of France, and the University of Hawaii.  This work is
based in part on data products produced at TERAPIX and the
Canadian Astronomy Data Centre (CADC) as part of the Canada-France-Hawaii
Telescope Legacy Survey, a collaborative project of NRC and CNRS.
We acknowledge the help of J.J. Kavelaars of the CADC Helpdesk about thumbnail
cutout for the ABC fields.
\\
This work is in part based on observations made with the
\textit{Spitzer Space Telescope}, which is operated by the Jet Propulsion Laboratory,
California Institute of Technology under NASA. Support for this work, part of
the \textit{Spitzer Space Telescope} Legacy Science Program, was provided by NASA
through an award issued by the Jet Propulsion Laboratory, California Institute of
Technology under NASA contract 1407. 
\\
This work is in part based on data collected within the UKIDSS survey.
The UKIDSS project uses the UKIRT Wide Field Camera funded by the UK Particle
Physics and Astronomy Research Council (PPARC).
Financial resources for WFCAM Science Archive development
were provided by the UK Science and Technology Facilities
Council (STFC; formerly by PPARC).
\\
\textit{GALEX} is a NASA mission managed by the Jet Propulsion Laboratory.
\textit{GALEX} data used in this paper were obtained from the Multimission Archive at the 
Space Telescope Science Institute (MAST). STScI is operated by the Association of 
Universities for Research in Astronomy, Inc., under NASA contract NAS5-26555. Support 
for MAST for non-HST data is provided by the NASA Office of Space Science via grant 
NNX09AF08G and by other grants and contracts.
\\
OM, AE and JS acknowledge support from the ESA PRODEX Programmes "XMM-LSS" and "XXL", 
and from the Belgian Federal Science Policy Office.
They also acknowledge support from the Communaut\'e fran\c{c}aise de Belgique -
Actions de recherche concert\'es - Acad\'emie universitaire Wallonie-Europe".

 \clearpage
\appendix
\section{Listings of database content}
\label{SecAPPTables}

\begin{table*}
\caption{List of parameters provided in the public \xmm-LSS
catalogues. All are available at the \xmm-LSS Milan database in the
separate tables {\tt 2XLSSB} or {\tt 2XLSSBd} for the soft band
and {\tt 2XLSSCD} or {\tt 2XLSSCDd} for
the hard band. The column name has an appropriate prefix: when
there are two column names given, one with the prefix B and one
with the prefix CD, only the one applicable to the given band
appears in the relevant table but both may show up in the
band-merged tables {\tt 2XLSS} or {\tt 2XLSSd}
(the family of tables  {\tt 2XLSSd}, {\tt 2XLSSBd} and  {\tt 2XLSSCDd} constitutes
the full exposure catalogue, while {\tt 2XLSS}, {\tt 2XLSSB} and  {\tt 2XLSSCD}
the 10 ks one);
column names without prefix are relevant to the
individual band only. The last four  columns indicate
respectively: (X)  whether a parameter is natively computed by
{\sc Xamin}; (m) whether a parameter is available also in the
band-merged table; (o) whether a parameter is present in the multiwavelength
table together with those described in Table
\ref{TabMulti}; and (C) whether a parameter is present in the
catalogue stored at CDS. \label{TabBand}}
\begin{tabular}{lllllll}
   \hline
   Column name & units & meaning and usage & X & m & o & C \\
   \hline\hline
{\tt Bseq     or CDseq    } & --    & internal sequence number (unique)                     &   & X & X & X \\
{\tt Bcatname or CDcatname} & --    & IAU catalogue name {\tt 2XLSS{\it x} Jhhmmss.s-ddmmss}, {\tt{\it x}=B or CD} &   & X & X & X \\
{\tt Xseq}                & --    & numeric pointer to merged entry see Table \ref{TabMerge}      &   & X & X & X \\
{\tt Xcatname}            & --    & name pointer to merged entry see Table \ref{TabMerge}      &   & X & X & X \\
{\tt Xlsspointer}         & --    & {\tt Xseq} of corresponding source in \xlss~ version I catalogue &   & X & X & X \\
{\tt Xdeep}               & --    & {\tt Xseq} of corresponding source in \xlssd~ (in table \xlssII~only) &   & X & X & X \\
{\tt Xfield}              & --    & \xmm~ pointing number (internal use)                  &   & X & X &   \\
{\tt FieldName}             & --    & \xmm~ pointing name                                   &   & X & X &   \\
{\tt Xbadfield}           & 0 or 1  & pointing is good (0) or bad (1)                     &   & X & X &   \\
{\tt expm1}               & s     & MOS1 camera exposure in the band                      & X &   &   &   \\
{\tt expm2}               & s     & MOS2 camera exposure in the band                      & X &   &   &   \\
{\tt exppn}               & s     & pn   camera exposure in the band                      & X &   &   &   \\
{\tt gapm1}               & \arcsec & MOS1 distance to nearest gap                        & X &   &   &   \\
{\tt gapm2}               & \arcsec & MOS2 distance to nearest gap                        & X &   &   &   \\
{\tt gappn}               & \arcsec & pn   distance to nearest gap                        & X &   &   &   \\
{\tt Bnearest or CDnearest} & \arcsec & distance to nearest detected neighbour              & X & X &   &   \\
{\tt Bc1c2}               & 0|1|2   & 1 for class C1, 2 for C2, 0 for undefined                 &   & X & X & X \\
{\tt CDc1c2}              & 0|1|2   & 1 for class C1, 2 for C2, 0 for undefined                 &   &   &   &   \\
{\tt Bcorerad or CDcorerad} & \arcsec & core radius EXT (for extended sources)              & X & X &   & X \\
{\tt Bextlike or CDextlike} & --      & extension likelihood EXT\_LH                        & X & X &   & X \\
{\tt Bdetlik\_pnt or CDdetlik\_pnt} & --      & detection likelihood DET\_LH for point-like fit      & X &   &   &   \\
{\tt Bdetlik\_ext or CDdetlik\_ext} & --      & detection likelihood DET\_LH for extended  fit      & X &   &   &   \\
{\tt Boffaxis or CDoffaxis} & \arcmin & off-axis angle                                      &   & X &   & X \\
{\tt Brawra\_pnt or CDrawra\_pnt}   & $\degr$ & source RA  (not astrometrically corrected) for point-like fit      & X &   &   &   \\
{\tt Brawdec\_pnt or CDrawdec\_pnt} & $\degr$ & source Dec (not astrometrically corrected) for point-like fit      & X &   &   &   \\
{\tt Brawra\_ext or CDrawra\_ext}   & $\degr$ & source RA  (not astrometrically corrected) for extended  fit      & X &   &   &   \\
{\tt Brawdec\_ext or CDrawdec\_ext} & $\degr$ & source Dec (not astrometrically corrected) for extended  fit      & X &   &   &   \\
{\tt Bra\_pnt or CDra\_pnt}         & $\degr$ & source RA  (astrometrically corrected) for point-like fit      & X &   &   &   \\
{\tt Bdec\_pnt or CDdec\_pnt}       & $\degr$ & source Dec (astrometrically corrected) for point-like fit      & X &   &   &   \\
{\tt Bra\_ext or CDra\_ext}         & $\degr$ & source RA  (astrometrically corrected) for extended  fit      & X &   &   &   \\
{\tt Bdec\_ext or CDdec\_ext}       & $\degr$ & source Dec (astrometrically corrected) for extended  fit      & X &   &   &   \\
{\tt Bposerr or CDposerr} & \arcsec & error on coordinates according to Table \ref{Tab8PapI}&   & X &   & X \\
{\tt Bratemos\_pnt or CDratemos\_pnt} & ct/s  & MOS count rate for point-like fit          & X &   &   &   \\
{\tt Bratepn\_pnt or CDratepn\_pnt}   & ct/s  & pn  count rate for point-like fit          & X &   &   &   \\
{\tt Bratemos\_ext or CDratemos\_ext} & ct/s  & MOS count rate for extended fit           & X &   &   &   \\
{\tt Bratepn\_ext or CDratepn\_ext}   & ct/s  & pn  count rate for extended fit           & X &   &   &   \\
{\tt countmos\_pnt  } & ct   & MOS number of counts for point-like fit          & X &   &   &   \\
{\tt countpn\_pnt   } & ct   & pn  number of counts for point-like fit          & X &   &   &   \\
{\tt countmos\_ext  } & ct   & MOS number of counts for extended fit           & X &   &   &   \\
{\tt countpn\_ext   } & ct   & pn  number of counts for extended fit           & X &   &   &   \\
{\tt bkgmos\_pnt    } & ct/pixel/detector & MOS local background for point-like fit          & X &   &   &   \\
{\tt bkgpn\_pnt     } & ct/pixel & pn  local background for point-like fit          & X &   &   &   \\
{\tt bkgmos\_ext    } & ct/pixel/detector & MOS local background for extended fit           & X &   &   &   \\
{\tt bkgpn\_ext     } & ct/pixel & pn  local background for extended fit           & X &   &   &   \\
{\tt Bflux or CDflux}                & erg/cm$^{2}$/s & source flux (undefined i.e. -1 for extended)  &   & X &   & X \\
{\tt Bfluxflag or CDfluxflag}        & 0 to 2         & 0 if MOS-pn difference $<20\%$, 1 between 20\%-50\%, 2 above 50\% &   & X &   & X \\
   \hline
\end{tabular}
\end{table*}

The list of columns in the X-ray single-band and band-merged tables is almost
the same as in Paper I, however is reported for completeness in
Tables \ref{TabBand} and \ref{TabMerge}.
As for Paper I, the {\it main}
parameters (as flagged in such tables)
of the merged X-ray  catalogue will be available in electronic form also
at the CDS. 
The multiwavelength tables have instead a substantially increased number of
columns, which are listed in Table~\ref{TabMulti}.

\begin{table*}
\caption{List of database parameters,  as Table  \ref{TabBand},
but for the additional columns present only in the merged
catalogue tables {\tt 2XLSS} or {\tt 2XLSSd}.  When there are two column names
given, one with the prefix B and one with the prefix CD, they
relate to the given band, and both show up in the band-merged
table. Column names with the prefix X are relevant to merged
properties. } \label{TabMerge}
\begin{tabular}{lllllll}
   \hline
   Column name & units & meaning and usage &X & m & o & C \\
   \hline\hline
{\tt Xseq}                  & --      & Internal sequence number (unique)                 &   & X & X & X \\
{\tt Xcatname}              & --      & IAU catalogue name {\tt 2XLSS{\it d} Jhhmmss.s-ddmmss{\it c}}, see Section \ref{SecNaming}    &   & X & X & X \\
{\tt Bspurious and CDspurious} & 0 or 1  & set to 1 when soft/hard component has DET\_LH$<15$ &   & X &   &   \\
{\tt Bdetlike and CDdetlike}   & --      & detection likelihood EXT\_LH (pnt or ext according to source class)           & X & X &   & X \\
{\tt Xra}                   & $\degr$ & source RA  (astrometrically corrected) (pnt or ext acc. to source class in best band) &   & X & X & X \\
{\tt Xdec}                  & $\degr$ & source Dec (astrometrically corrected) (pnt or ext acc. to source class in best band) &   & X & X & X \\
{\tt Bra and CDra}           & $\degr$ & source RA  (astrometrically corrected) (pnt or ext according to source class) &   & X & X & X \\
{\tt Bdec and CDdec}         & $\degr$ & source Dec (astrometrically corrected) (pnt or ext according to source class) &   & X & X & X \\
{\tt Xbestband}             & 2 or 3  & band with highest likelihood: 2 for B, 3 for CD  &   & X &   &   \\
{\tt Xastrocorr}            & 4 or 5  & astrometric correction from CFHTLS (4) or USNO (5) or none (0) &   & X &   &   \\
{\tt Xmaxdist}              & \arcsec & distance between B and CD positions               &   & X &   &   \\
{\tt Xlink}                 & --      & pointer to Xseq of secondary association, see Section \ref{SecNaming}  &   & X &   &   \\
{\tt Bratemos and CDratemos}   & ct/s/detector   & MOS count rate (pnt or ext according to source class)  & X & X &   & X \\
{\tt Bratepn and CDratepn}     & ct/s   & pn  count rate (pnt or ext according to source class)  & X & X &   & X \\
   \hline
\end{tabular}
\end{table*}

\begin{table*}
\caption{ List of additional non-X-ray 
columns in the multiwavelength {\tt 2XLSSOPT} or {\tt 2XLSSOPTd} tables: they
also include columns with the X, B or CD prefixes from Tables \ref{TabBand} and 
\ref{TabMerge}, while those with the O, S, U or G prefix refer to optical, SWIRE,
UKIDSS or \textit{GALEX} data, and those without prefix  to combined properties.
\newline
 }
\label{TabMulti}
\begin{tabular}{lll}
   \hline
   Column name & units & meaning and usage \\
   \hline\hline
{\tt Xrank}    & -1 0 1 2  & ranking of the counterpart set (see \ref{SecOptProb}) \\
{\tt Ou}       & magnitude &  $u^*$ magnitude             \\
{\tt Og}       & magnitude &  $g'$ magnitude              \\
{\tt Or\_}       & magnitude &  $r'$ magnitude            \\
{\tt Oi}       & magnitude &  $i'$ magnitude              \\
{\tt Oz}       & magnitude &  $z'$ magnitude              \\
{\tt Ou\_e}    & magnitude &  error on $u^*$ magnitude    \\
{\tt Og\_e}    & magnitude &  error on $g'$ magnitude     \\
{\tt Or\_e}    & magnitude &  error on $r'$ magnitude     \\
{\tt Oi\_e}    & magnitude &  error on $i'$ magnitude     \\
{\tt Oz\_e}    & magnitude &  error on $z'$ magnitude     \\
{\tt OseqD1}   & --        &  internal sequence number (if magnitudes from D1 field) \\
{\tt OseqW1}   & --        &  internal sequence number (in W1 or ABC fields)         \\
{\tt Ofield}   & --        &  CFHT  field identification in form $\pm x \pm y$ or D1 or A,B,C (see \ref{SecOptInput} and  Fig.~\ref{FigPointing}) \\ 
{\tt Ora}      & $\degr$   &  RA of the optical candidate           \\
{\tt Odec}     & $\degr$   &  Declination of the optical candidate  \\
{\tt Oflag}    & --        &  binary flag combining 0/1 galaxy/star, 0/4 normal/masked, 0/8 normal/saturated  \\
{\tt Sf36}     & $\mu$Jy   &  IRAC flux at 3.6 $\mu$m     \\
{\tt Sf45}     & $\mu$Jy   &  IRAC flux at 4.5 $\mu$m     \\
{\tt Sf58}     & $\mu$Jy   &  IRAC flux at 5.8 $\mu$m     \\
{\tt Sf80}     & $\mu$Jy   &  IRAC flux at 8.0 $\mu$m     \\
{\tt Sf24}     & $\mu$Jy   &  MIPS flux at 24  $\mu$m     \\
{\tt Sf70}     & $\mu$Jy   &  MIPS flux at 70  $\mu$m     \\
{\tt Sf160}    & $\mu$Jy   &  MIPS flux at 160 $\mu$m     \\
{\tt Sf36\_e}  & $\mu$Jy   &  error on 3.6 $\mu$m flux    \\
{\tt Sf45\_e}  & $\mu$Jy   &  error on 4.5 $\mu$m flux    \\
{\tt Sf58\_e}  & $\mu$Jy   &  error on 5.8 $\mu$m flux    \\
{\tt Sf80\_e}  & $\mu$Jy   &  error on 8.0 $\mu$m flux    \\
{\tt Sf24\_e}  & $\mu$Jy   &  error on 24 $\mu$m flux     \\
{\tt Sf70\_e}  & $\mu$Jy   &  error on 70 $\mu$m flux     \\
{\tt Sf160\_e} & $\mu$Jy   &  error on 160 $\mu$m flux    \\
{\tt Sseq}     & --        &  internal sequence number for SWIRE objects    \\
{\tt Sra}      & $\degr$   &  RA of the SWIRE   candidate           \\
{\tt Sdec}     & $\degr$   &  Declination of the SWIRE   candidate  \\
{\tt Uj}       & magnitude &  J magnitude             \\
{\tt Uh}       & magnitude &  H magnitude             \\
{\tt Uk}       & magnitude &  K magnitude             \\
{\tt Uj\_e}    & magnitude &  error on J magnitude    \\
{\tt Uh\_e}    & magnitude &  error on H magnitude    \\
{\tt Uk\_e}    & magnitude &  error on K magnitude    \\
{\tt Useq}     & --        &  internal sequence number for UKIDSS objects    \\
{\tt Ura}      & $\degr$   &  RA of the UKIDSS   candidate           \\
{\tt Udec}     & $\degr$   &  Declination of the UKIDSS   candidate  \\
{\tt Gnuv}     &     mJy   &  \textit{GALEX} near UV flux          \\
{\tt Gfuv}     &     mJy   &  \textit{GALEX}  far UV flux          \\
{\tt Gnuv\_e}  &     mJy   &  error on near UV    flux    \\
{\tt Gfuv\_e}  &     mJy   &  error on  far UV    flux    \\
{\tt Gseq}     & --        &  internal sequence number for \textit{GALEX} objects    \\
{\tt Gra}      & $\degr$   &  RA of the \textit{GALEX}   candidate           \\
{\tt Gdec}     & $\degr$   &  Declination of the \textit{GALEX}   candidate  \\
{\tt distXO}   & \arcsec   &  distance from the X-ray corrected position to the optical position \\
{\tt distXS}   & \arcsec   &  distance from the X-ray corrected position to the SWIRE   position \\
{\tt distXU}   & \arcsec   &  distance from the X-ray corrected position to the UKIDSS  position \\
{\tt distXG}   & \arcsec   &  distance from the X-ray corrected position to the \textit{GALEX}   position \\
{\tt probXO}   & --        &  chance probability of X-ray to optical association (see \ref{SecOptProb})      \\
{\tt probXS}   & --        &  chance probability of X-ray to SWIRE   association (see \ref{SecOptProb})      \\
{\tt probXU}   & --        &  chance probability of X-ray to UKIDSS  association (see \ref{SecOptProb})      \\
{\tt probXG}   & --        &  chance probability of X-ray to \textit{GALEX}   association (see \ref{SecOptProb})      \\
   \hline
\end{tabular}
\end{table*}

\section{Details of probability computation}
\label{SecAPPOptProb}

\begin{table*}
\caption{Parameters used for probability computation.}
 \label{TabProb} 
         \begin{tabular}{lllrrl}
         \hline
Probability  & $m$ & density $n(brighter~than~m)$ & a        & b         & Notes   \\
         \hline\hline
$probXO$     & $i'$        & $n(<i')=10^{a+b i'}$ & -9.32415 & 0.293833  & for D1 field \\
             &             &                      & -9.23183 & 0.290519  & for W1 excluding ABC fields\\
             & $r'$        & $n(<r')=10^{a+b r'}$ & -9.18619 & 0.279706  & for W1 ABC fields\\
$probXS$     & $F_\lambda$ & $n(>F_\lambda)=10^{a+b*log(F_\lambda)}$ & & & in increasing order of $\lambda$ whichever first\\
             & $\lambda = 3.6\mu m$ &             & -1.68062 & -0.944191 & \\
             & $\lambda = 4.5\mu m$ &             & -1.73693 & -0.976644 & \\
             & $\lambda = 5.8\mu m$ &             & -2.04933 & -0.829700 & \\
             & $\lambda = 8.0\mu m$ &             & -1.49944 & -1.07201  & \\
             & $\lambda = 24 \mu m$ &             & 0.102480 & -1.53410  & \\
$probXU$     & $J$         & $n(<J)=10^{a+bJ}$    & -8.67503 & 0.268272  & taken best if both bands are present \\
             & $K$         & $n(<K)=10^{a+bK}$    & -8.96264 & 0.321560  & \\
$probXG$     & $NUV$       & $n(<NUV)=10^{a+bNUV}$  & -11.0875 & 0.326965  & taken best if both bands are present \\
             & $FUV$       & $n(<FUV)=10^{a+bFUV}$  & -13.9827 & 0.433838  & \\
\hline
\end{tabular}
\end{table*}

Probabilities of chance association between a counterpart and an X-ray source are
computed using formula (\ref{Eq1}) in Section~\ref{SecOptProb},
where 
the density $n(brighter~than~m)$ is computed from simple
linear fits as reported in Table \ref{TabProb}. The same table indicates also the
magnitudes or fluxes used to look-up at the density for the appropriate band.
We did only a rough estimate of the parametrization of the density for the
entire \xmm-LSS area, neglecting any possible spatial variation.

X-ray to CFHTLS probability, called ${\tt probXO}$, is computed for sources with a
CFHTLS counterpart in the order: D1 if present, else W1.
In the case of undefined CFHTLS magnitudes, the pointing tile limiting magnitude
was used (read directly from the W1 table, or fixed to $i'=25$ for D1).

The other probabilities, X-ray to SWIRE probability ${\tt probXS}$,
X-ray to UKIDSS probability ${\tt probXU}$, and X-ray to GALEX probability ${\tt probXG}$,
are computed as given in the notes in Table \ref{TabProb}.

A probability of 99 ("undefined") is assigned whenever it could not be computed.

\begin{figure*}
   \centering
   \includegraphics[width=168mm]{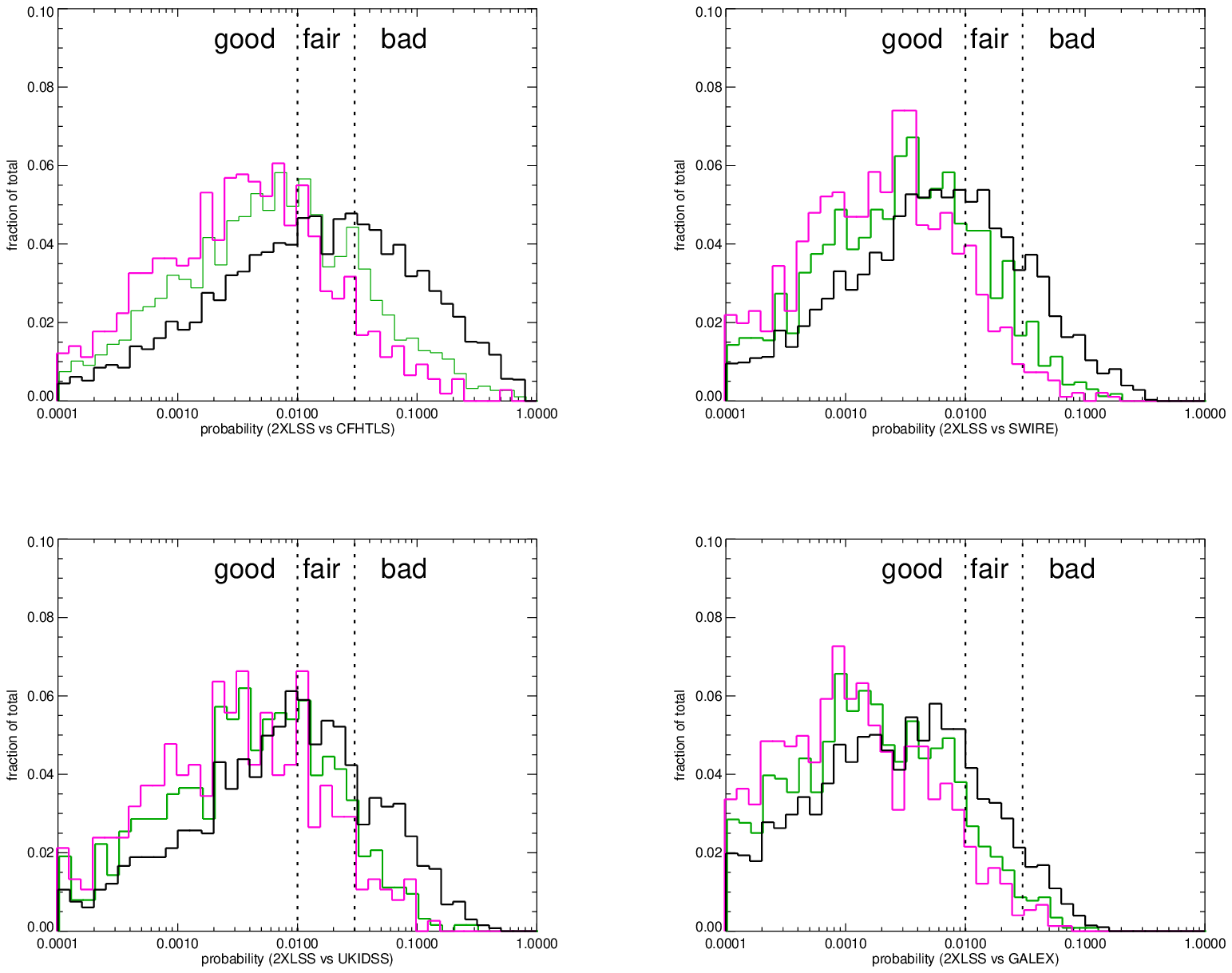} 
   \caption{Histograms of the four probabilities (${\tt probXO}$, ${\tt probXS}$,
            ${\tt probXU}$ and ${\tt probXG}$) normalized to the total number
            of best counterparts without any undefined probability in the
            total sample (thick black), with a detection likelihood of at least 40 ($3 \sigma$)
            in the best band (thin light gray, green on the web version), or of at least 75 ($4 \sigma$, thick dark gray, magenta on the web version).
            The dashed fiducial lines identify the loci with good, fair,
            or bad probability as defined in Section~\ref{SecOptProb}.
             }
    \label{FigHisto}
    \end{figure*}

A statistics of the probability ranking defined in Section~\ref{SecOptProb}
is shown in Fig.~\ref{FigHisto}.
This figure indicates an acceptable tuning with the CFHTLS, SWIRE and
UKIDSS catalogues, while \textit{GALEX} data are perhaps \textit{overtuned} in the 
sense that there is an excess of good probabilities. This may indicate that the probability
computation has to be revised.

\section{Comparison between catalogues}
\label{SecAPPComp}

We provide here details on the comparisons: 
(a) between \xlssd~ and Version I \xlss~ (Section~\ref{SecXstatNewOld});
(b) between {\tt 2XLSSOPT} and {\tt 2XLSSOPTd} (Section \ref{SecAPPCompOct});
(c) between our catalogues and the XMDS one (Section \ref{SecAPPXMDS});
(d) between our catalogues and the 2XMM one (Section \ref{SecAPPWatson}).

   \begin{figure*}
   \centering
   \includegraphics[width=8.5cm]{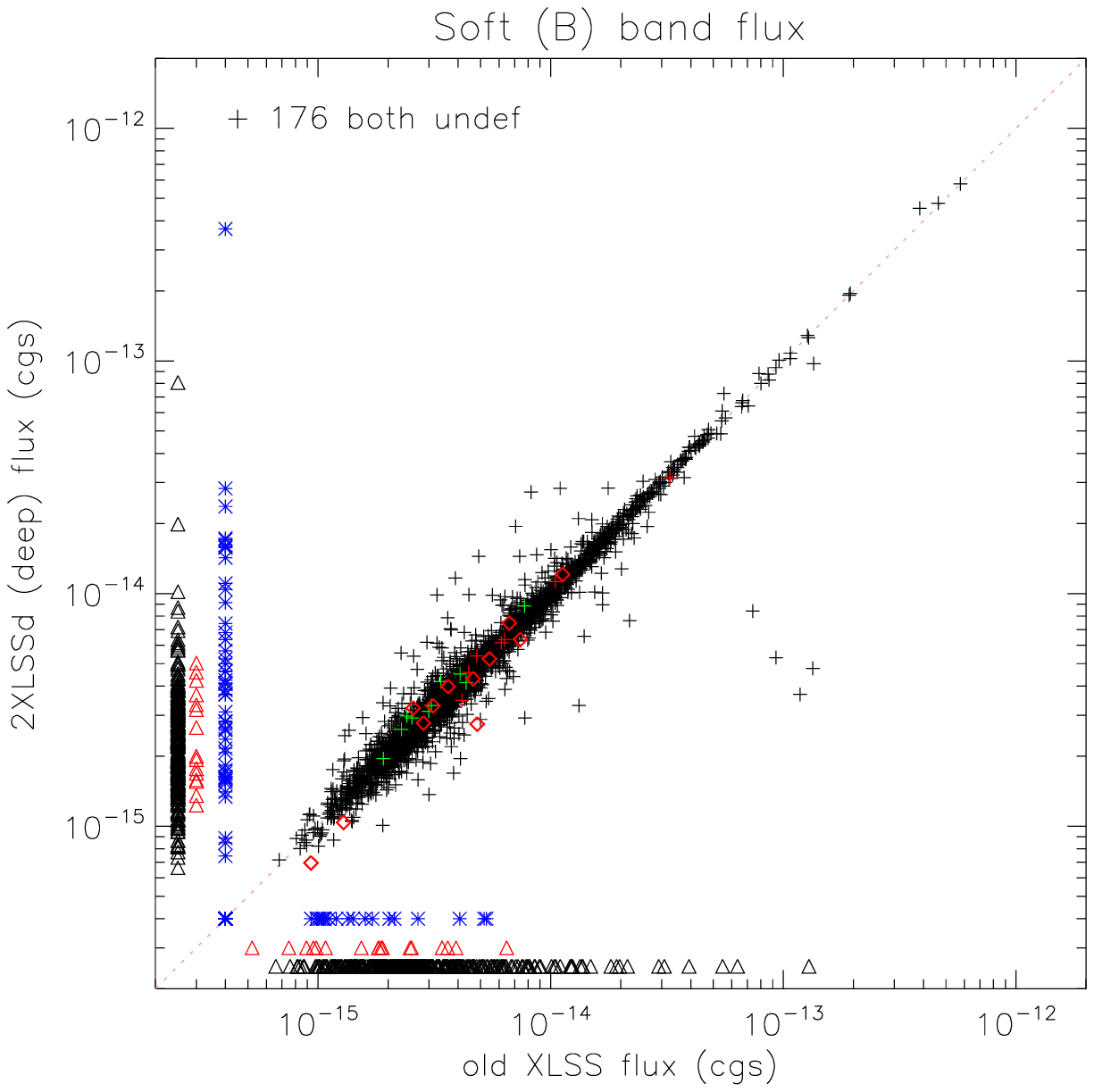} 
   \includegraphics[width=8.5cm]{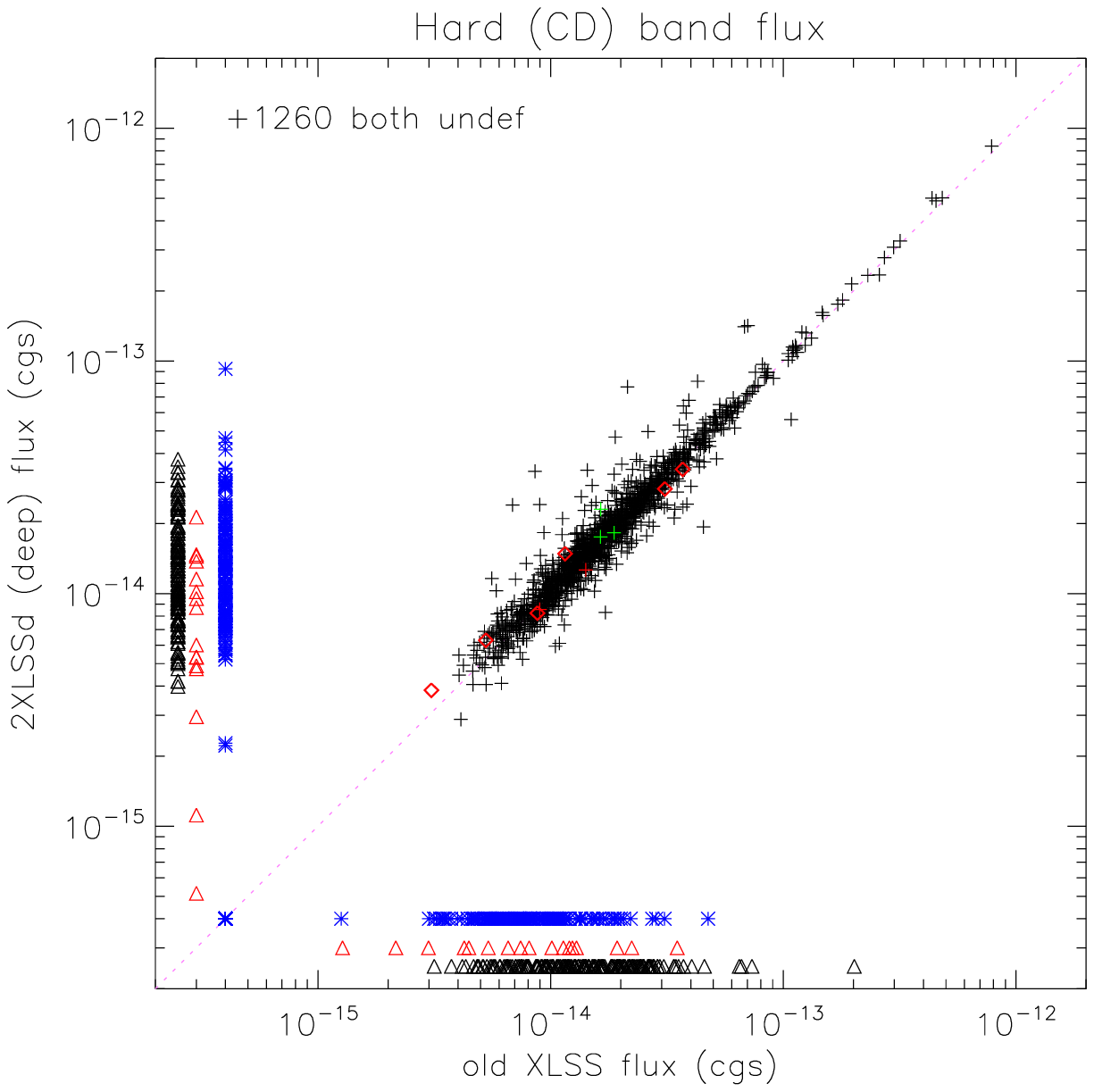} 
   \caption{Comparison of
            the flux 
            in the soft (left column) and hard (right column) energy bands
            between the Version I \xlss~ catalogue and \xlssd.
            Symbols are identical to those of Fig.~\ref{FigCompa2}.
             }
   \label{FigCompa3}
    \end{figure*}

\subsection{Comparison between \xlss~ and \xlssd}  
\label{SecXstatNewOld}

We first checked
that the new pipeline version provides results consistent with the previous IDL
version by performing detailed tests on simulated and real \xmm~ pointings. 
The detection parts of both pipelines give nearly identical results for 
point-like and extended sources. The characterization parts (maximum 
likelihood fitting) are in excellent agreement for point-like sources. 
Regarding extended sources, comparison of fluxes, sizes and likelihoods 
estimates from the two pipelines are in very good agreement. Some 
differences do however show up for individual faint sources or sources 
close to the detector borders and gaps. These differences can be 
attributed to statistical fluctuations, and comparing these values to the 
input characteristics of simulated sources in a statistical sense, we 
conclude that both pipelines perform equivalently. 
Finally we compared directly the results.

The association between sources in the Version 2 catalogues (\xlssII~ and \xlssd)
with the earlier \xlss~ release (in the smaller area covered by the latter)
is possible using the database column ${\tt Xlsspointer}$, as explained in
Section~\ref{SecNaming}.
2824 sources are common between \xlss~ and \xlssd~ (out of 3385 in \xlss).
There are 452 objects appearing only in \xlssd~ in the pointings covered by
\xlss, and 561 \xlss~ objects not confirmed in \xlssd. A majority (respectively
95\% of the former and 88\% of the latter) have a poor detection likelihood
($LH<40$ corresponding conventionally to $<3\sigma$).
As an example of the rather good agreement between the two catalogue versions
(mainly between the older and newer \xamin~
version, and also between the 6\arcsec~ vs 10\arcsec~ band merging)
we compare the 
fluxes between \xlss~
and \xlssd~ (which makes sense since \xlss~ was also using full exposures), via
the plots given in Fig.~\ref{FigCompa3}.

\subsection{Comparison between {\tt 2XLSSOPT} and {\tt 2XLSSOPTd} \label{SecAPPCompOct}} 

There are 20837 potential counterpart sets for 6721 X-ray sources in input for {\tt 2XLSSOPTd}, and
16813 for 5572 sources for {\tt 2XLSSOPT} (the comparison includes negative rank \textit{rejected}
counterparts since rejection may act differently because of the displacement).
Of these, 1188 entries
(for 428 X-ray sources) in {\tt 2XLSSOPT} have no obvious correspondent in the deep
{\tt 2XLSSOPTd}.

On the other hand 15625 counterpart sets are associated with 5144 X-ray sources in
\xlssII~ with a corresponding source in \xlssd. These are the X-ray sources present
in both catalogues.

\begin{description}
\item
13454 counterpart sets (86\% of the common ones) for 4979 \xlssII~ sources (97\% of
the common ones) are identical (i.e. have the same counterparts in all non-X-ray catalogues).
 \begin{description}
 \item Of these, 139 are confirmed "blank fields" (no \textit{catalogued} counterpart in any waveband).
 \item Of the remaining 13315, 11556 (87\%) have the same rank in {\tt 2XLSSOPTd} and
       {\tt 2XLSSOPT}, namely, with reference to the rank definitions in Section \ref{SecOptProb}:
  \begin{description}
   \item 3710 are primary counterparts, 
   \item 1439 are secondary counterparts, 
   \item 6407 are rejected counterparts, not included in the catalogues.
   \vskip 0.1cm
   \item 1759 (of the 13315) have the same counterpart but with a different rank:
   \item for 511 of them the rank change is irrelevant (i.e. they remain anyhow the
         primary counterpart in both catalogues);
   \item 201 and 175 counterpart sets rejected in {\tt 2XLSSOPTd} are respectively 
         primary and secondary choices in {\tt 2XLSSOPT};
   \item 170 and 196 primary or secondary in {\tt 2XLSSOPTd} are rejected in {\tt 2XLSSOPT};
   \item 254 {\tt 2XLSSOPTd} primaries become secondary in {\tt 2XLSSOPT} while 252 undergo
         the opposite change from secondary to primary.
  \end{description}
 \end{description}
\item
In 38 additional cases (only 0.2\% of the common ones) the counterpart set
is the same between a couple of {\tt 2XLSSOPT} and {\tt 2XLSSOPTd} entries, but the latter are not
associated by column ${\tt Xdeep}$.
I.e. two sources have the same counterpart set
but are not the closest. In 10 cases this is due to ambiguous band merging, but in
the rest (which is anyhow an extremely small number) this probably means that there are
two 10 ks sources both displaced from but close to a given full exposure one (or v.v.).  
\item
The 2171 remaining {\tt 2XLSSOPT} counterpart sets (covering
1479 individual \xlssII~ X-ray sources) can be \textit{altogether different}
from all counterpart sets in {\tt 2XLSSOPTd} for the associated X-ray source (the
X-ray source position moved so much that entire counterpart sets are farther than
6\arcsec~ from either position), or may partially match (from 1 to 4 out of the 5 D1, W1,
SWIRE, UKIDSS or GALEX catalogues).
\item A breakdown of the partial matches (with a total larger than 2171,
because one specific counterpart set is compared with all the {\tt 2XLSSOPTd} counterpart
sets for the corresponding X-ray source) is:
 \begin{description}
  \item 2150 are no matches,
  \item  234 cases match in 1 catalogue,
  \item   66 cases match in 2 catalogues,
  \item   25 in 3,
  \item    7 in 4.
 \end{description}
\item An alternative breakdown (totalling 2171) is the following:
 \begin{description}
  \item 261 single no-match cases,
  \item 1601 multiple no-matches per X-ray source,
  \item 15 cases with one counterpart set with 1-3 matches,
  \item 51 cases with one no-match and one 1-3 matches,
  \item  6 cases with mixed matches,
  \item 237 with more no-match and mixed matches.
 \end{description}
\end{description}

A different approach for the comparison
is to consider only the \textit{best counterparts} i.e. those
ranked 0-1 (by definition one per X-ray source). Let us exclude the 1607 {\tt 2XLSSOPTd}
(24\% of 6721) X-ray sources not confirmed in the 10ks catalogue, and the 528
{\tt 2XLSSOPT} (8\% of 5572) new in the 10ks catalogue, and let us concentrate on
the common sources. Of their best counterparts:
\begin{description}
 \item $\approx 86$\% are essentially confirmed in both catalogues, namely
 \begin{description}
  \item 3\% are  confirmed tentative blank fields, 
  \item 72\% have the same counterparts and the same rank,
  \item 10\% have the same counterparts and compatible ranks, 
  \item only 1.4\% have partially matching counterpart sets with the same or compatible best rank.
 \end{description}
 \item A further 5\% and 4\% have the same counterparts but they are ranked differently (the
  best counterpart in one catalogue is either secondary or rejected in the other).
 \item The remaining 5\% have altogether different or partially matching counterparts which
  are ranked differently.
\end{description}
\textit{So the difference between counterparts in the deep and 10ks catalogues is
confined to less than 15\% of the common sources.}

\begin{figure*}
 \includegraphics[width=8cm]{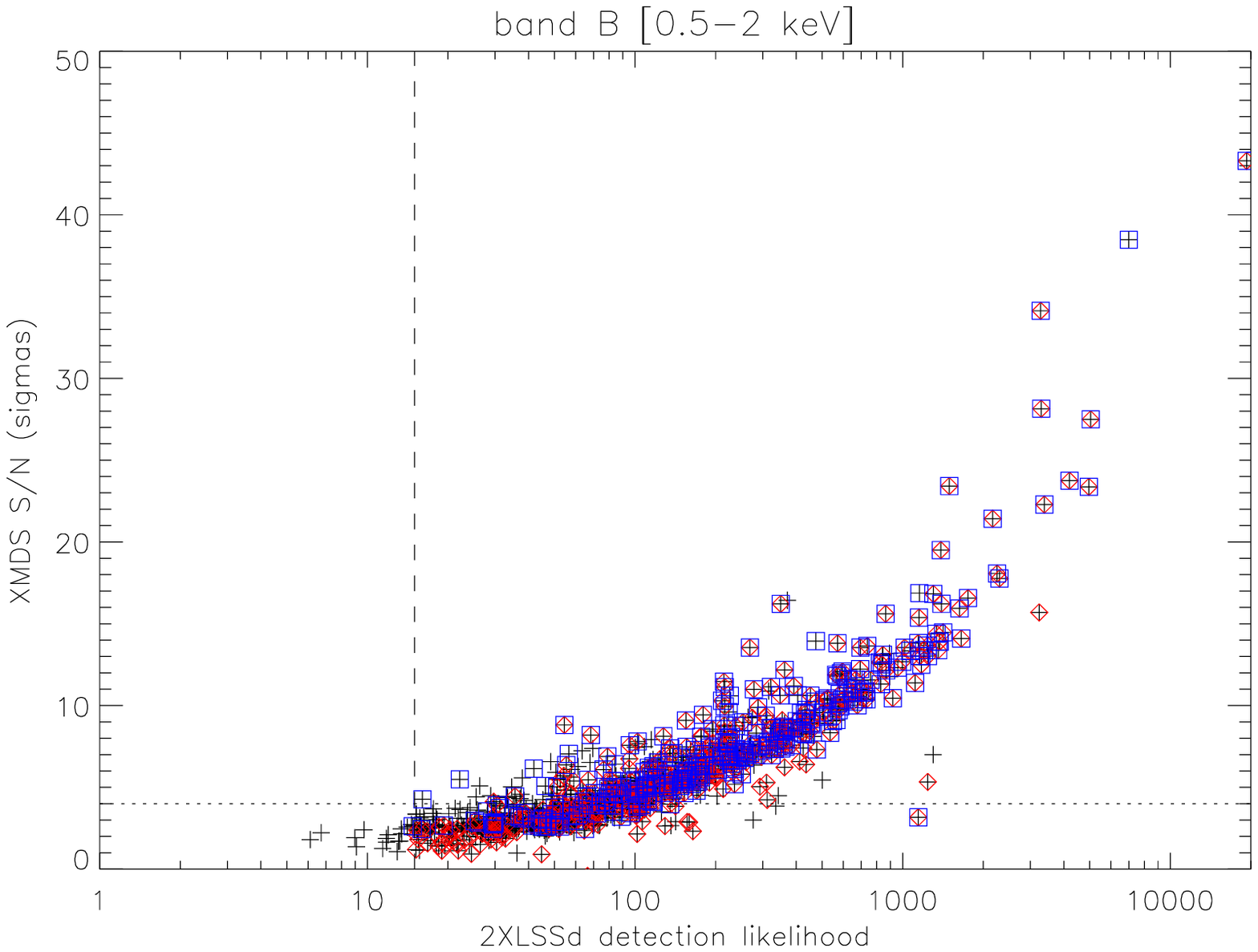} 
 \includegraphics[width=8cm]{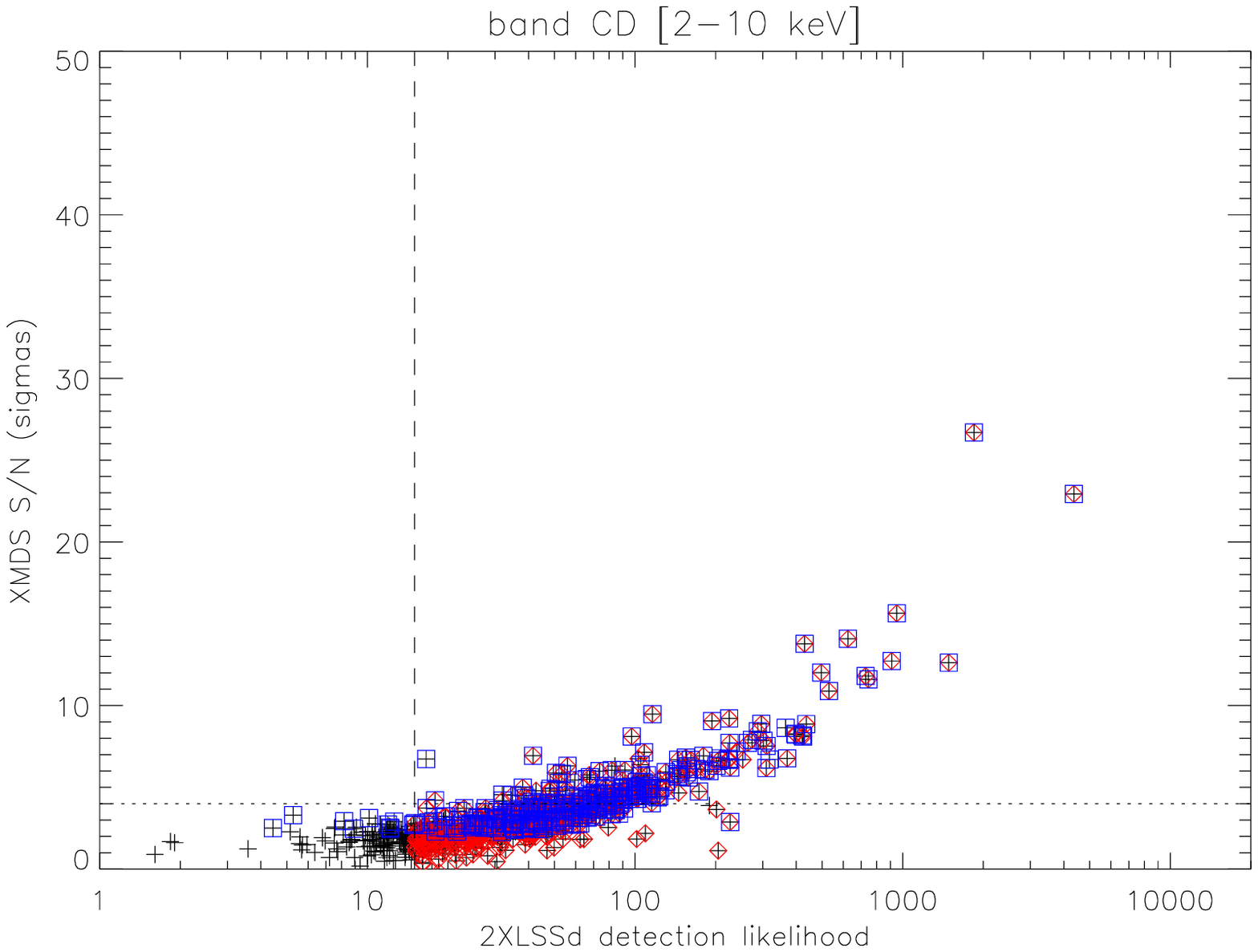} 
\caption{Cross calibration between the \xlssd~ (\xamin) detection likelihood and the
XMDS signal to noise ratio. Left panel for the soft band, right panel for
the hard band. The dashed vertical line indicates the \xlssd~ acceptance thresholds of
$LH>15$, while
the dotted horizontal line shows the conventional level of $4\sigma$.
Crosses indicate all objects detected in the given band.
A (red) diamond surrounds the sources detected above LH threshold {\it in both bands}
in \xlssd, while a (blue) square surrounds those detected above the chance probability threshold {\it in both bands}
in the XMDS.
}
         \label{xmdssn}
\end{figure*}
\begin{figure*}
 \includegraphics[width=8cm]{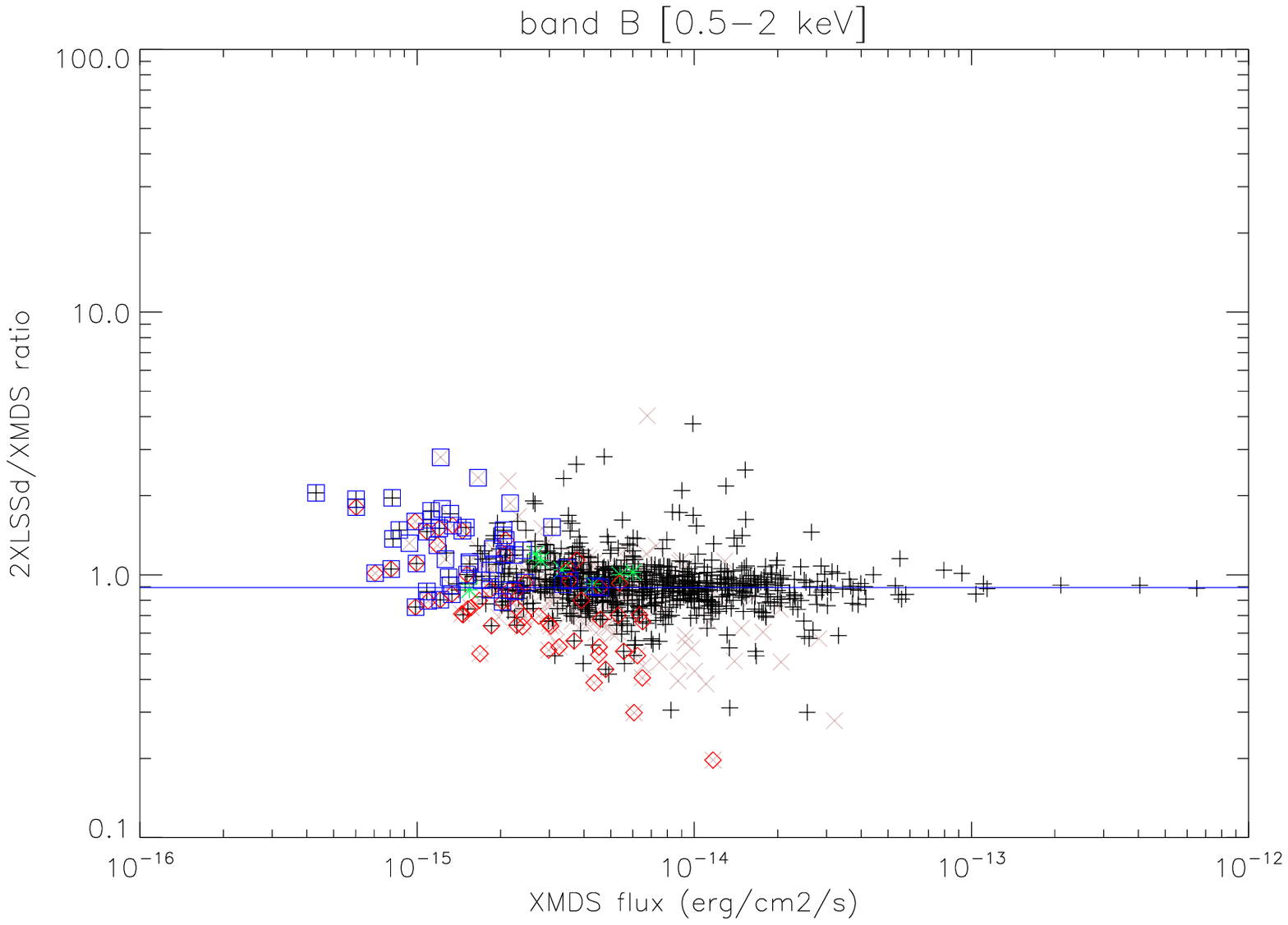} 
 \includegraphics[width=8cm]{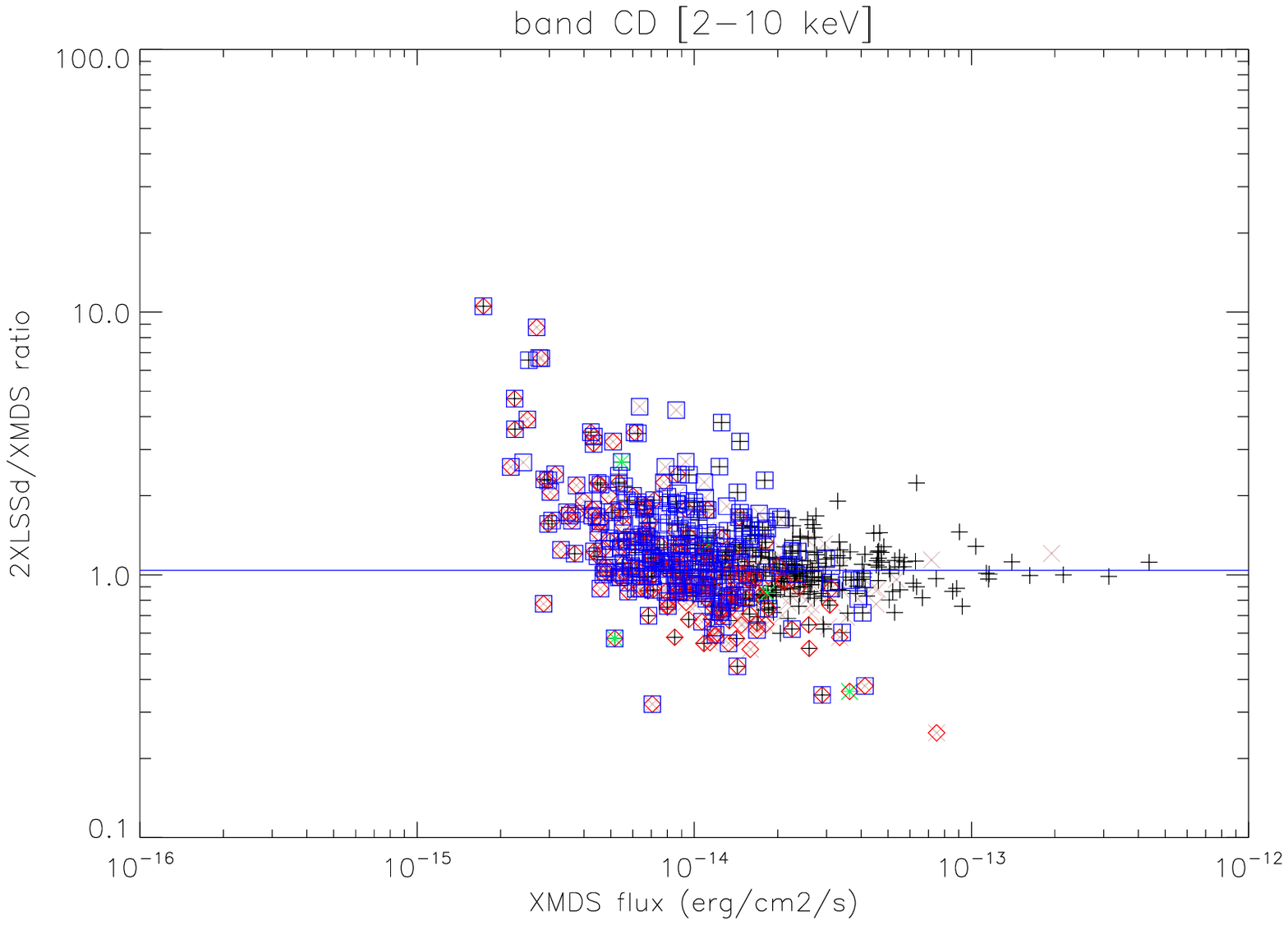} 
\caption{
The ratio of the \xlssd~ and XMDS fluxes as a function
of the XMDS flux for band B (left panel) and band CD (right panel).
The horizontal solid line is a fiducial line
corresponding to the actual average ratio in the band (see text). 
The (black) crosses indicate point-like sources which
have a ${\tt fluxflag}$ of 0 or 1, the (pink) X those with a {\tt fluxflag} of 2,
i.e. where the MOS and pn fluxes differ by more than 50\%. The (green) asterisks
correspond to extended C2 sources for which the flux is computed from the point-like
rates (C1 sources have flux set to undefined and are not plotted).
A (red) diamond surrounds the points with a poor \xlssd~ likelihood $15<LH<20$ in the band.
A (blue) square surrounds the points with a poor XMDS probability $prob>2\times10^{-4}$ in the band.
Note that the latter symbols are different from those used in Fig. \ref{xmdssn}.
 }
         \label{xmdsflux}
\end{figure*}

\subsection{Comparison with XMDS} 
\label{SecAPPXMDS}

A subset of the XMDS (\xmm~Medium Deep Survey) data was published as
the XMDS/VVDS $4\sigma$ catalogue (\citealp{2005A&A...439..413C}; the same paper
contained also the logN-LogS of the entire catalogue).
Contextually to the release of the entire XMDS catalogue through our
database contextually with Version 2 \xmm-LSS ones (see Table~\ref{TabTables}),
we report here some details of
a comparison between the catalogues presented in this paper
(mainly \xlssd) and XMDS.

The main procedural differences between the two pipelines can be summarized as follows:

\begin{enumerate}
\item The XMDS covers the fields G01 to G19 in Table~\ref{TabPointing}
\item XMDS data were analysed with a \textit{fully independent} (and more
     traditional) pipeline based on the one by \cite{2002ApJ...564..190B}.
\item The XMDS pipeline uses the {\sc SAS} to detect candidates in 5 energy bands 
      simultaneously (and not in 2 independent bands with later merging)
      operating on event files merged
      from all 3 \xmm~cameras and from the entire \xmm~field of view
      (not just the central 13\arcmin),
\item It then applies the \cite{2002ApJ...564..190B} characterization,
      which is robust but oriented to point sources only (unlike the wavelet method in \xamin~
      which handles better extended sources).
\item The event pattern selection is different (non-standard and broader in XMDS).
\item The removal of redundant sources is handled differently, in particular
      the primary detection is chosen differently, the position is inherited from the
      primary detection, but the flux is obtained stacking data from all overlapping
      pointings. 
\item The astrometric correction offsets are different.
\item Also the XMDS catalogue does not include spurious objects (but
      only those above a probability threshold), so the
      difference between the raw database table and the catalogue is only due to the
      overlap removal procedure.
\end{enumerate}

\subsubsection{Comparison of the X-ray source lists \label{SecAPPXMDSX}}  
~\newline

As anticipated in Section~\ref{SecXMDS},
the XMDS catalogue includes 1168 sources, by definition all in the G-labelled fields.
Comparison with the \xlssd~ or \xlssII~ catalogues may also involve
adjacent B fields if the \xmm-LSS overlap removal procedure preferred those.
The association between
XMDS and \xmm-LSS sources is done within a radius of 10\arcsec.
The comparison occurs naturally with the deep catalogue, because it involves
the \textit{same} input (ODF) event data (for the full exposures) with different pipelines and procedures.
Of those 1168 objects:
\begin{description}
\item 1082 have a counterpart in the full exposure input tables,
\item of which 1057 are catalogued in \xlssd;
\item while 1019 have one in the 10 ks input tables,
\item of which 956 in \xlssII. 
\end{description}

Of the 86 XMDS sources not in the input table for \xlssd:
\begin{description}
\item 23 are at off-axis angles greater than 13\arcmin~ (ignored by \xamin), 
\item and 39 anyhow at large off-axis angles ($>10\arcmin$),
so it is not surprising they were excluded by \xamin;
\item
similarly 43 are potential ultrasoft sources (the band with the
highest S/N ratio in XMDS is the A band [0.3-0.5 keV], which is not processed
by the current release of \xamin), so they are legitimately excluded;
\item
considering the XMDS significance, 46 and 72 are
respectively below $3\sigma$ and $4\sigma$;
\item
if one allows the combination of different conditions, a net majority of
the XMDS-only sources (76) are \textit{either} ultrasoft,
or at offaxis $>10\arcmin$ or at $<3\sigma$.
\end{description}

We concentrate below on the 1057 XMDS sources with a \xlssd~
correspondent (which can be called \textit{common catalogued}). Of them:
\begin{description}
\item 783 are in the same (G) field, so should be exactly the same detections;
\item 173 are stacked XMDS entries;
\item the remaining 103 cases associate sources detected in different pointings:
 39 in another G field, 64 in a B field.
\end{description}

Concerning the distance between the (astrometrically corrected) X-ray positions in the 
XMDS and \xlssd~ catalogues, 58\% of the sources are closer than
2\arcsec, 88\% closer than 4\arcsec~ and only 4\% more distant than 6\arcsec,
in general concentrated among the sources with lesser significance, and the few
extended ones.
The agreement between XMDS and \xlssd~ positions, peaking around 1\arcsec, is
better than the typical inter-band distance between \xlssd~ detections in the two
energy bands, which peaks around 2\arcsec.  
Compare panels (d) and (c) of Fig.~\ref{FigDistHisto}.

We have cross calibrated graphically the detection likelihood of
\xamin~ with the chance probability of the XMDS (for definition see \citealt{2002ApJ...564..190B})
and the detection likelihood of
\xamin~ with the significance in terms of number of $\sigma$ of the XMDS (see
\cite{2005A&A...439..413C} and references therein). We only show the latter
in Fig. \ref{xmdssn}.
One can see that a likelihood of 75 corresponds more or less to the $4\sigma$ level,
and one of 40 to the $3\sigma$ level.

We also note that 89\% of the common sources have B as the best band (highest likelihood)
in \xlssd. To be more precise:
\begin{description}
\item 96\% of the sources are observed by \xlssd~ in the B band,
\item 62\% are observed in the CD band, and
\item 57\% are observed in both.
\end{description}
This can be compared with the totality of the \xlssd~ catalogue, where
(with no appreciable difference between the full \xlssd~
catalogue and the sources in the G fields alone):
\begin{description}
\item 84\% of the sources have B as the best band,
\item 89\% are observed in  the B band,
\item 48\% in the CD band, and
\item 37\% in both.
\end{description}
The XMDS by construction includes measurements in all 5 energy bands even if the source 
is above the probability threshold only in one. If we consider as good detections for XMDS
only those with $prob<2\times10^{-4}$ in the band, we have that,
 of the sources in common with \xlssd:
\begin{description}
\item 91\% are detected in the B band,
\item 36\% in the CD band, and
\item 30\% in both.
\end{description}

The fluxes, computed for XMDS according to the prescriptions of \cite{2002ApJ...564..190B} and
for \xmm-LSS as explained in Section~\ref{SecFlux}, are compared in Fig. \ref{xmdsflux}.
Extended sources classified C1 are excluded as their \xlssd~ flux is set to undefined.
The fluxes match qualitatively, although there is a systematic difference: namely
\xlssd~ fluxes are 
0.895 lower than the XMDS fluxes in the B band, while they are only
1.040 higher in the CD band. 

It can be seen that a larger scatter in fluxes occurs for the sources which have
poorer significance in either catalogue, while outliers are generally due to sources
presumably falling near a chip gap on one detector (and as such characterized by
a {\tt fluxflag} of 2), or exceptionally by residual C2 extended sources (for which
the \xmm-LSS flux is computed from the point-like rate).

\subsubsection{Comparison of the optical counterparts \label{SecAPPXMDSOpt}} 
~\newline

It is also possible, similarly to what was done in Section \ref{SecCompOct}, to compare the
counterparts in optical (and other) bands between the XMDS catalogue and one of
our \xmm-LSS catalogues. In doing this one should consider:

\begin{enumerate}
\item that the identification procedure for XMDS is historically
      different from the one used here, in particular it was done in several
      incremental steps, and uses distances capped to 2\arcsec~ in the
      computation of  probabilities,
\item that the catalogues used for XMDS were in larger number (for a total
      of 27) and included many other, including older,
      data sources (e.g. VVDS, radio data, SIMBAD and NED, CFHTLS T003, etc.)
\end{enumerate}

Therefore we present a comparison limited to the {\tt 2XLSSOPTd} catalogue (the one which
matches better XMDS in exposures), and, for XMDS, to \textit{reduced counterpart sets}
considering only CFHTLS T004 D1 and W1, SWIRE DR6 and GALEX. UKIDSS was not
included since  {\tt 2XLSSOPTd} uses release DR5, while XMDS used release DR3.
More specifically we consider only the 1057 X-ray sources in common between
XMDS and \xlssd, as described in \ref{SecAPPXMDSX}.

They correspond to 4316 counterpart sets (of any rank)
in {\tt 2XLSSOPTd} and 4916 in XMDS.
We found that a large fraction (3620) of the possible
counterpart sets are identical (i.e. they have the same counterparts,
irrespective of ranking, in both catalogues). 

   \begin{figure*}
   \centering
   \includegraphics[width=8.5cm]{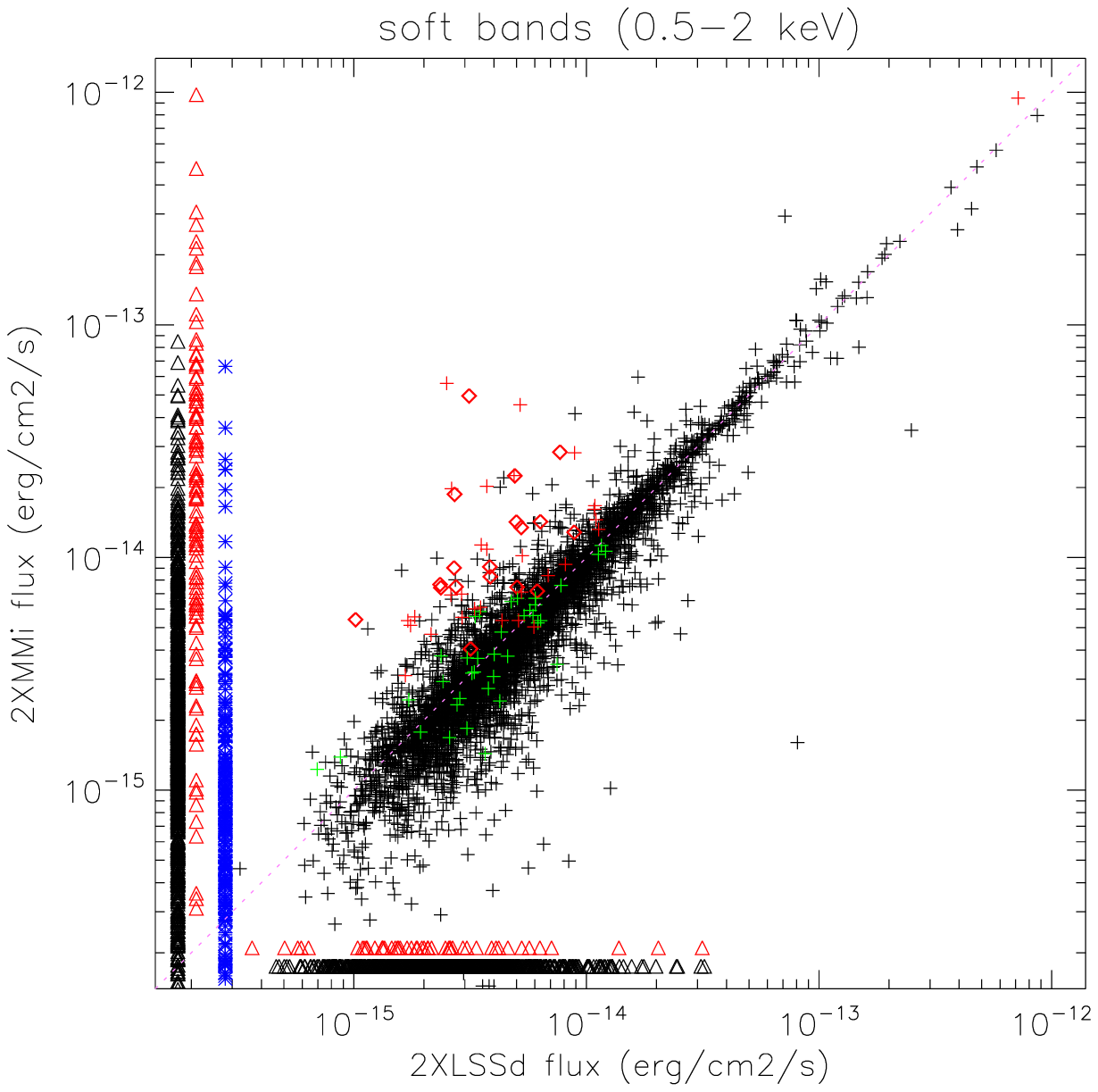} 
   \includegraphics[width=8.5cm]{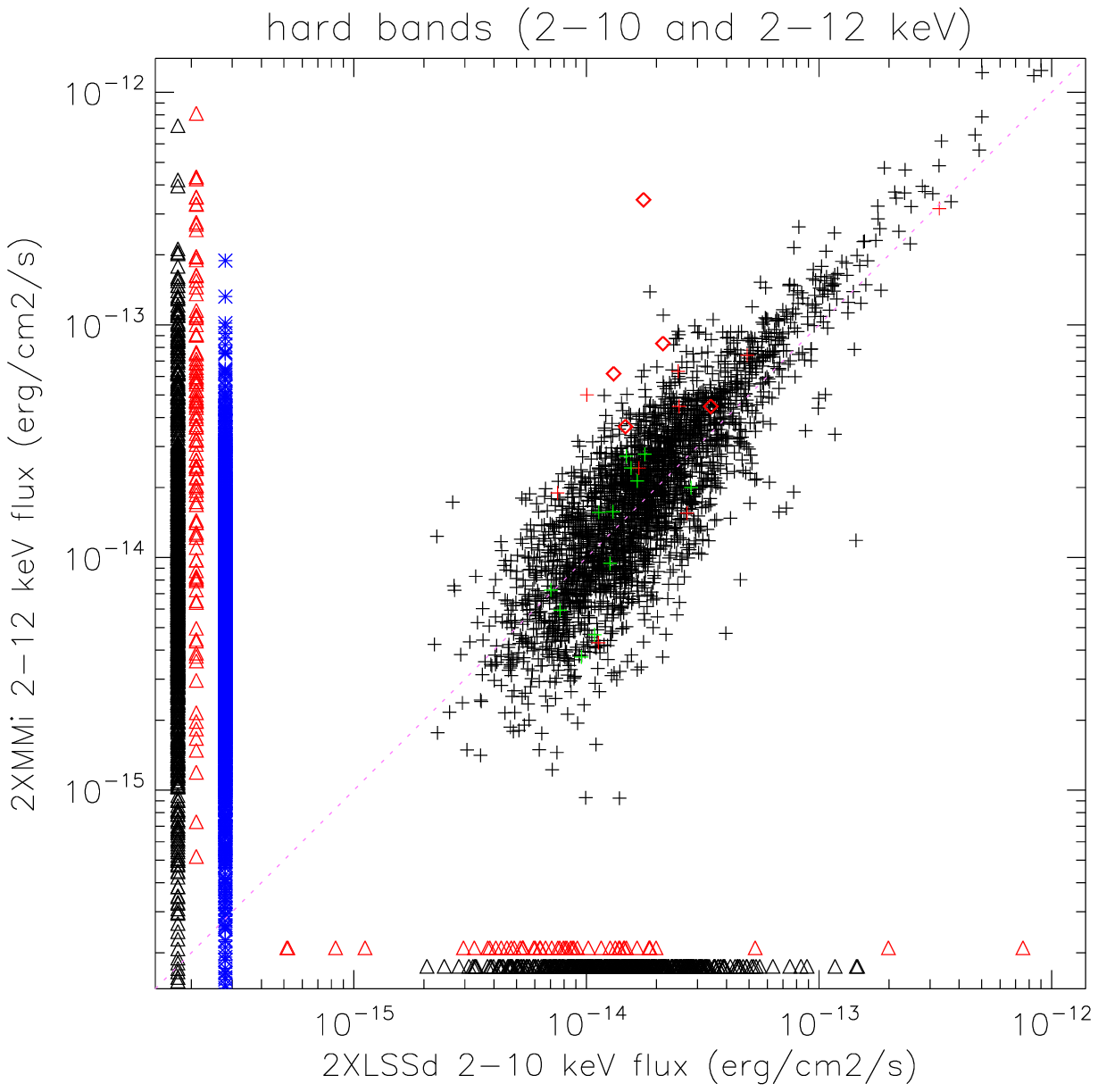} 
   \caption{Comparison of the flux
            in the soft (left panel) and hard (right panel) energy bands.
            between \xlssd~ and 2XMM.
            Crosses and diamonds indicate point-like or extended objects
            associated in the two catalogues (see text).
            Blue asterisks indicate fluxes present but
            \textit{undefined} in \xlssd,
            while triangles indicate sources present only in one catalogue
            (both are placed at a conventional out-of-range X or Y coordinate).
            Colour coding (only in the web version) is as follows:
            black cross for point-like common sources,
            red diamond for extended sources in both \xlssd~ and 2XMM,
            green cross for \xlssd~ extended
            object point-like in 2XMM; vice versa for red cross;
            triangles are black or red for point-like or extended sources.
            Remember that fluxes for C1 extended sources are undefined in
            our database (see Table 9 in Paper I).
             }
   \label{Fig2XMM}
    \end{figure*}

We can then concentrate on the \textit{best counterparts} (ranks 0-1; a similar
ranking system, though different in details, was used also for XMDS):
\begin{description}
\item
for 627 cases (59\% of 1057) the best counterpart is exactly the same in all 4
D1, W1, SWIRE and GALEX catalogues,
\item for 16\% in 3 catalogues (the other may be different or missing),
\item for 15\% in 2,
\item for 4\% in 1;
\item making a total of 81\% with the choice of a highly compatible counterpart.
\item
A very limited number of cases (5 and 8) are potential "blank fields"
respectively in {\tt 2XLSSOPTd} and XMDS, with another counterpart in the
other catalogue.
\item
The remaining 193 (18\%) cases select an altogether different counterpart
in the two catalogues:
 \begin{description}
 \item for 97 {\tt 2XLSSOPTd} and 77 XMDS sources
  the counterpart set is present only in one catalogue and fully replaced by
  something else in the other;
 \item in other 96 or 116 cases, the counterpart set which is preferred in one catalogue 
 is still present in the other with a different rank (secondary or rejected). 
 \end{description}
\end{description}

In conclusion, the compatibility between the counterparts is satisfactory.

\subsection{Comparison with 2XMM} 
\label{SecAPPWatson}

We quickly compared our \xlssd~ catalogue with the 2XMM  catalogue 
\citep{2009A&A...493..339W}. Namely we used the 2XMMi-DR3 "slim" reduced 
catalogue\footnote{\url{http://xmmssc-www.star.le.ac.uk/Catalogue/2XMMi-DR3cat_slim_v1.0.csv.gz}},
which contains exposure-merged sources (not individual detections), and thus somewhat
compare with our post overlap-removal catalogues.

We restricted a comparison to a rectangular area fully encompassing \xlssd~ and
noticed that such an area includes precisely just our pointings, with the only exception of a
single additional pointing centered on the bright star Mira Ceti 
(ObsId 014850~0201). 

The rectangular area contains 6181 2XMM sources. Since the slim catalogue does not 
contain indication on the pointings, we can tentatively flag sources in the Mira Ceti
field as the 60 within 13\arcmin~ from the respective pointing centre. We checked the
association with our 6721 \xlssd~ sources within the customary radius of 10\arcsec. 
We find:
\begin{description}
\item 5039 sources are associated
\item 1682 \xlssd~ ones are not associated
\item 1141 2XMM ones are not associated
\end{description}
We note that 59\% of the unassociated \xlssd~ sources have a rather poor likelihood
($ML<20$), and 93\% are below $ML<40$ (i.e. $3\sigma$).
Despite the different definition of the 2XMM likelihood (parameter
${\tt SC\_DET\_ML}$), we note that of 1141 unassociated 2XMM sources, 67\% have
${\tt SC\_DET\_ML}<15$.

Of the associated cases, all but 69 have a single association. The ambiguous ones
are all plain couples, and usually well separated (one 2XMM source at 1-2\arcsec~
from the \xlssd~ position, and the other at 8-9\arcsec).

The distance between the astrometrically corrected position (2XMM used the USNO
catalogue for this purpose) for the associated primary (closest) cases is within 
2\arcsec~ in 55\% of the cases, within 4\arcsec~ in 86\% and within 6\arcsec~ in 95\% 
(while 90\% of the 69 secondaries are above 6\arcsec). The histogram of distances 
however peaks at a 1\arcsec~ offset (not unlike XMDS vs \xlssd).

The common subset contains 100 of our extended sources, and 88 2XMM extended sources
(i.e. those with their parameter ${\tt SC\_EXTENT}>0$). 56 are considered extended in
both catalogues.
110 of our extended sources do not appear at all in 2XMM, and 113 2XMM extended
sources do not appear in \xlssd.

The 2XMM slim catalogue does not contain count rates. It contains fluxes in several
energy bands, but these are different from ours. We can directly compare the sum
of 2XMM bands 2 and 3 with our B band (0.5-2 keV), while at harder energies we can
compare the sum of 2XMM bands 4 and 5 (2-12 keV) with our CD band (2-10 keV).

The comparison of the fluxes is reported in Fig.~\ref{Fig2XMM}. The average 
2XMM/\xlssd~ flux ratio for common point-like sources
is 0.92 in the soft band (same energy range), and
1.22 in the hard bands (where 2XMM extends 2 keV further). Again, considering
the differences in the pipelines, the agreement is acceptable.

 \bsp
 \label{lastpage}
\end{document}